\documentclass{ws-mpla}

\usepackage[super]{cite}
\usepackage{xcolor}
\usepackage[verbose,hypertexnames=false]{hyperref}
\hypersetup{colorlinks=false,allbordercolors=blue,pdfborderstyle={/S/U/W 1}}

\usepackage{amsmath}
\usepackage{amssymb}
\usepackage{graphicx}

\begin{document}


\markboth{Djedai Ayang Kamo et al.}
{}

\catchline{}{}{}{}{}

\title{Thermodynamics, Joule-Thomson Expansion and Ruppeiner Geometry of a Dyonic NED-AdS Black Hole Surrounded by Quintessence\\
}

\author{
	Djedai Ayang Kamo$^{1,\ast}$,
	Saleh Mahamat$^{2}$,
	Ragil Brand Tsafack Ndongmo$^{3}$,
	Kamiko Kouemeni Jean Rodrigue$^{1}$,
	Thomas Bouetou Bouetou$^{5}$,
	Timoleon Crepin Kofane$^{6}$
}

\address{
	$^{1}$Department of Physics, Faculty of Science, University of Maroua,\\
	P.O. Box 814, Maroua, Cameroon
}

\address{
	$^{2}$Research Center in Didactics of Fundamental and Applied Sciences (CERDISFA),\\
	Department of Physics, Higher Teachers' Training College,\\
	University of Maroua, P.O. Box 55, Maroua, Cameroon			
}

\address{
	$^{3}$Department of Physics, Faculty of Science, University of Yaound\'e I,\\
	P.O. Box 812, Yaound\'e, Cameroon
}

\address{
	$^{4}$National Advanced School of Engineering, University of Yaound\'e I,\\
	P.O. Box 8390, Yaound\'e, Cameroon
}

\address{
	$^{5}$Department of Physics and Astronomy,\\
	Botswana International University of Science and Technology,\\
	Private Mail Bag 16, Palapye, Botswana
}

\maketitle

\begin{history}
\received{(Day Month Year)}
\revised{(Day Month Year)}
\accepted{(Day Month Year)}
\published{(Day Month Year)}
\end{history}

\begin{abstract}
We investigate the thermodynamic properties, Joule-Thomson expansion, and microscopic structure of a dyonic Reissner-Nordstr\"om-AdS black hole in nonlinear electrodynamics surrounded by quintessence. Within the extended phase-space formalism, analytical expressions for the mass, Hawking temperature, equation of state, heat capacity, Gibbs free energy, Joule-Thomson coefficient, and Ruppeiner curvature are derived. The effects of the nonlinear-electrodynamics parameter $\omega$, the quintessence parameters $(c,\epsilon)$, the dyonic charge $Q$, and the pressure $P$ are systematically investigated. The temperature profiles and equation of state exhibit a Van der Waals--like phase structure, with critical points strongly modified by the quintessence and nonlinear electrodynamivs contributions. The heat capacity and Gibbs free energy further reveal the corresponding thermodynamic stability and phase-transition behaviour. The Joule-Thomson analysis shows distinct cooling and heating regions separated by inversion points. In particular, quintessence generally lowers the critical and inversion temperatures, while increasing the charge and varying the nonlinear parameter significantly deform the inversion structure. The Ruppeiner curvature exhibits pronounced peaks and sign changes near critical regions, indicating strong thermodynamic correlations and alternating attractive and repulsive effective microscopic interactions. The limiting cases $\omega\sim 1$, $c\sim 0$, and $Q\sim 0$ consistently recover the corresponding Maxwell, no-quintessence, and neutral black hole behaviours. These results demonstrate that the combined effects of nonlinear electrodynamics, dyonic charge, quintessence, and AdS pressure substantially reshape both the macroscopic phase structure and the effective microscopic interactions of the black hole.
\end{abstract}

\ccode{PACS Nos.: 03.65.$-$w, 04.62.+v}

\section{Introduction}
\label{sec:introduction}
The discovery that the Universe is undergoing accelerated expansion has profoundly influenced our understanding of cosmology and gravitation. This accelerated expansion is generally attributed to a mysterious form of energy known as dark energy, which dominates the large-scale dynamics of the Universe and remains one of the major challenges in modern physics [1-3]. These developments have stimulated increasing interest in the interplay between cosmological phenomena and compact astrophysical objects such as black holes. In particular, black holes can be regarded as thermodynamic systems characterized by entropy, temperature, and pressure, and can exhibit phase transitions analogous to those observed in conventional thermodynamic systems [4-9].
Within the framework of extended black hole thermodynamics, the cosmological constant $\Lambda$ is identified with the thermodynamic pressure, $P=-\frac{\Lambda}{8\pi},$ while the black-hole mass $M$ is interpreted as the enthalpy of the system rather than its internal energy [10,11]. This interpretation provides a natural framework for investigating black-hole phase transitions and critical phenomena. In particular, charged AdS black holes exhibit thermodynamic behaviour analogous to that of Van der Waals fluids, including first- and second-order phase transitions, coexistence regions, and critical points [12].
Among the simplest charged black-hole configurations, the Reissner-Nordstr\"om (RN) solution provides a useful framework for studying the thermodynamic properties of charged gravitational systems. When both electric and magnetic charges are present, the resulting configuration is referred to as a dyonic black hole and exhibits richer electromagnetic and thermodynamic properties [13-16]. However, the standard RN solution is based on Maxwell's linear electrodynamics, whose applicability may become limited in strong-field regimes. Nonlinear electrodynamics (NED), originally introduced in the Born-Infeld framework to improve the behaviour of electromagnetic fields in strong-field regimes [17], provides a natural extension of Maxwell theory. NED black hole solutions have been extensively investigated, revealing significant modifications of the electromagnetic field, thermodynamic quantities, and stability properties [18-22].
More recently, NED black holes in asymptotically anti-de Sitter spacetimes have attracted considerable attention, including in the context of the AdS/CFT correspondence [23-27]. These studies have shown that nonlinear electromagnetic corrections can substantially modify the critical temperature, pressure, and global phase structure of black holes. Moreover, the nonlinear electromagnetic sector can provide additional information about the microscopic properties of black-hole systems, particularly when combined with thermodynamic-geometrical approaches such as Ruppeiner geometry.
In parallel, cosmological observations have motivated the incorporation of dark-energy models into black-hole spacetimes. Among the possible descriptions of dark energy, quintessence provides a simple and effective phenomenological framework characterized by an parameter satisfying $-1<\epsilon<-\frac{1}{3},$ [28,29]. The presence of quintessence modifies the spacetime structure and introduces an additional contribution to the gravitational potential, thereby affecting the thermodynamic properties of black holes [30-34]. Previous studies have demonstrated that quintessence can significantly modify phase transitions and critical phenomena in charged and rotating black hole systems [35-38]. The above range of $\epsilon$ corresponds to a fluid with negative pressure and provides a physically relevant description of quintessence. Consequently, the simultaneous presence of quintessence and nonlinear electrodynamics offers an interesting framework for investigating the interplay between strong field electromagnetic effects and cosmological dark energy backgrounds in AdS black hole thermodynamics.
Within this framework, the Joule-Thomson (JT) expansion provides an important tool for investigating the thermal response of black holes during isenthalpic processes. The JT coefficient characterizes the variation of temperature with pressure at constant enthalpy and allows one to identify the inversion points separating the cooling and heating regimes. The JT expansion has therefore been widely employed to investigate the effects of charge, nonlinear electromagnetic corrections, and dark energy on the thermodynamic behaviour of black holes [39-49].
Complementary to the macroscopic thermodynamic analysis, thermodynamic geometry, and in particular Ruppeiner geometry, provides a geometrical framework for investigating the effective microscopic interactions of black hole systems [50-65]. The Ruppeiner metric is defined on the space of equilibrium thermodynamic states, while its associated scalar curvature can provide information about the effective microscopic correlations of the underlying system. In the conventional interpretation, negative and positive values of the Ruppeiner curvature are associated with predominantly attractive and repulsive effective interactions, respectively, while divergences of the curvature may signal critical behaviour and phase-transition phenomena.
Despite the considerable progress achieved in the individual study of nonlinear electrodynamics, dyonic black holes, quintessence, Joule-Thomson expansion, and thermodynamic geometry, their combined effects remain comparatively less explored. In particular, the simultaneous presence of nonlinear electromagnetic corrections, dyonic charge, quintessence, and extended phase space thermodynamics may generate nontrivial modifications of the critical and microscopic properties of the system. Nonlinear electrodynamics modifies the electromagnetic contribution to the black hole geometry, while the parameter $\omega$ controls the departure from the linear Maxwell regime. The dyonic configuration incorporates both electric and magnetic charge contributions, whereas quintessence introduces an additional cosmological contribution associated with negative pressure. Investigating the interplay among these effects in an AdS background is therefore important for obtaining a more complete understanding of the thermodynamic and microscopic structure of charged black holes.
The aim of the present work is twofold. First, we investigate the thermodynamic properties and Joule-Thomson expansion of a dyonic Reissner-Nordstr\"om-AdS black hole modified by nonlinear electrodynamics and surrounded by quintessence. We derive the corresponding expressions for the black hole mass, Hawking temperature, entropy, thermodynamic volume, equation of state, heat capacity, and Gibbs free energy. The associated critical points and phase-transition behaviour are then examined by systematically analysing the effects of the nonlinear electrodynamics parameter $\omega$, the quintessence parameters $(c,\epsilon)$, the dyonic charge $Q$, and the thermodynamic pressure $P$. We further determine the Joule-Thomson coefficient and the corresponding inversion curves in order to characterize the cooling and heating regions of the black hole.
Second, we investigate the microscopic structure of the system using Ruppeiner thermodynamic geometry. By constructing the thermodynamic metric and evaluating its scalar curvature, we analyse the effective microscopic interactions and their dependence on the black-hole and model parameters. This allows us to establish a connection between the macroscopic thermodynamic behaviour and the effective microscopic structure of the dyonic NED black hole surrounded by quintessence.
The paper is organized as follows. Section~2 introduces the black hole solution and develops its thermodynamic properties. Section~3 is devoted to the Joule-Thomson expansion, including the JT coefficient and inversion curves. Section~4 investigates the microscopic structure through Ruppeiner thermodynamic geometry. Finally, Section~5 summarizes the main results and discusses their physical implications.
	
\section{Black Hole Solution and Thermodynamics}
\label{sec:thermodynamics}
We consider a static and spherically symmetric dyonic black hole in an asymptotically anti-de Sitter spacetime, modified by nonlinear electrodynamics (NED) and surrounded by quintessence. The nonlinear electromagnetic sector is described by the conformal- and $SO(2)$-duality-invariant NED model introduced in Ref.~\cite{Amirabi2021}. In the extended phase space, the cosmological constant is interpreted as the thermodynamic pressure.

\subsection{Dyonic NED black hole solution surrounded by quintessence}

The gravitational and nonlinear electromagnetic dynamics are described by the action
\begin{equation}
\mathcal{I}
=
\frac{1}{16\pi}
\int d^4x\,\sqrt{-g}
\left[
R-2\Lambda+\mathcal{L}(\mathcal{F},\mathcal{G})
\right]
+\mathcal{I}_{q}
+\mathcal{I}_{\rm bdy},
\label{eq:action}
\end{equation}
where $R$ is the Ricci scalar, $\Lambda$ is the cosmological constant, $\mathcal{I}_{\rm bdy}$ denotes the appropriate boundary contribution, and $\mathcal{I}_{q}$ represents the phenomenological quintessence sector. The electromagnetic invariants are defined as
\begin{equation}
\mathcal{F}=F_{\mu\nu}F^{\mu\nu},
\qquad
\mathcal{G}=F_{\mu\nu}\widetilde{F}^{\mu\nu},
\label{eq:invariants}
\end{equation}
where $\widetilde{F}^{\mu\nu}$ is the dual electromagnetic tensor.

Following Ref.~\cite{Amirabi2021}, we adopt the nonlinear
electrodynamics Lagrangian
\begin{equation}
\mathcal{L}(\mathcal{F},\mathcal{G})
=
-\mathcal{F}\cosh\gamma
+
\sqrt{\mathcal{F}^{2}+\mathcal{G}^{2}}\,
\sinh\gamma ,
\label{eq:NED_Lagrangian}
\end{equation}
where $\gamma$ is the nonlinear-electrodynamics parameter.
The causal and unitary structure of this theory requires
\begin{equation}
0<\gamma<
\gamma_{\rm max},
\qquad
\gamma_{\rm max}
=
\tanh^{-1}\left(\frac{\sqrt{2}}{2}\right).
\label{eq:gamma_bound}
\end{equation}
In the limit $\gamma\sim 0$, one obtains
$\mathcal{L}\sim-\mathcal{F}$, recovering the linear Maxwell theory.

For the dyonic configuration, we consider the metric
\begin{equation}
ds^{2}
=
-f(r)\,dt^{2}
+\frac{dr^{2}}{f(r)}
+r^{2}
\left(
d\theta^{2}
+\sin^{2}\theta\,d\phi^{2}
\right),
\label{eq:metric_ansatz}
\end{equation}
and the electromagnetic two-form
\begin{equation}
F
=
E(r)\,dt\wedge dr
+
Q_m\sin\theta\,d\theta\wedge d\phi ,
\label{eq:dyonic_field}
\end{equation}
where $Q_e$ and $Q_m$ denote the electric and magnetic charges,
respectively.

The nonlinear Maxwell equations together with the Bianchi identity lead to the radial electric and magnetic fields
\begin{equation}
E(r)=\frac{Q_e}{r^{2}},
\qquad
B(r)=\frac{Q_m}{r^{2}}.
\label{eq:dyonic_fields}
\end{equation}
Consequently, the electromagnetic invariants become
\begin{equation}
\mathcal{F}
=
\frac{2(Q_m^{2}-Q_e^{2})}{r^{4}},
\qquad
\mathcal{G}
=
\frac{4Q_eQ_m}{r^{4}}.
\label{eq:EM_invariants}
\end{equation}

It is useful to introduce the charge ratio
\begin{equation}
q\equiv\frac{Q_e}{Q_m},
\label{eq:q_ratio}
\end{equation}
and the total dyonic charge
\begin{equation}
Q^{2}=Q_e^{2}+Q_m^{2}.
\label{eq:total_charge}
\end{equation}
For the nonlinear electrodynamics model~\eqref{eq:NED_Lagrangian}, the electromagnetic contribution to the energy-momentum tensor can be expressed in terms of the effective parameter $\omega$ defined by
\begin{equation}
\omega^{2}
=
\cosh\gamma\left(1-\frac{1-q^{2}}{1+q^{2}}\tanh\gamma\right).
\label{eq:omega_definition}
\end{equation}

With this definition, the nonlinear-electromagnetic energy-momentum tensor takes the compact form
\begin{equation}
T^{\mu}_{\ \nu\,({\rm NED})}
=
\frac{\omega^{2}Q^{2}}{r^{4}}
\,{\rm diag}(-1,-1,1,1).
\label{eq:NED_tensor}
\end{equation}

To incorporate the dark-energy environment, we adopt the phenomenological Kiselev description of quintessence~\cite{Kiselev2003}. Its energy-momentum
tensor is taken as
\begin{equation}
T^{\mu}_{\ \nu\,(q)}
=
{\rm diag}
\left[
-\rho_q(r),
-\rho_q(r),
\frac{1}{2}(3\epsilon+1)\rho_q(r),
\frac{1}{2}(3\epsilon+1)\rho_q(r)
\right],
\label{eq:quint_tensor}
\end{equation}
with
\begin{equation}
\rho_q(r)
=
\frac{c}{r^{3\epsilon+1}},
\label{eq:quint_density}
\end{equation}
where $c$ measures the strength of the quintessence contribution and $\epsilon$ is its equation-of-state parameter. The quintessence range is
\begin{equation}
-1<\epsilon<-\frac{1}{3}.
\label{eq:epsilon_range}
\end{equation}

The Einstein equations are
\begin{equation}
G_{\mu\nu}+\Lambda g_{\mu\nu}
=
8\pi
\left(
T_{\mu\nu}^{({\rm NED})}
+
T_{\mu\nu}^{(q)}
\right).
\label{eq:Einstein_equations}
\end{equation}

Using Eqs.~\eqref{eq:NED_tensor} and
\eqref{eq:quint_tensor} in the static and spherically symmetric
geometry~\eqref{eq:metric_ansatz}, the resulting metric function is
\begin{equation}
f(r)
=
1-\frac{2M}{r}
+\frac{\omega^{2}Q^{2}}{r^{2}}
-\frac{c}{r^{3\epsilon+1}}
-\frac{\Lambda r^{2}}{3}.
\label{eq:metric_function}
\end{equation}

The term $\omega^{2}Q^{2}/r^{2}$ represents the effective dyonic
electromagnetic contribution generated by the nonlinear electrodynamics.
The parameter $\omega$ is therefore not introduced as an independent phenomenological correction, but is determined by the underlying NED parameters $\gamma$ and $Q_e/Q_m$ through Eq.~\eqref{eq:omega_definition}.

We now adopt the extended phase space interpretation, in which the cosmological constant is regarded as a thermodynamic pressure,
\begin{equation}
P=-\frac{\Lambda}{8\pi}.
\label{eq:pressure}
\end{equation}
Thus,
\begin{equation}
\Lambda=-8\pi P,
\end{equation}
and the metric function can equivalently be written as
\begin{equation}
f(r)
=
1-\frac{2M}{r}
+\frac{\omega^{2}Q^{2}}{r^{2}}
-\frac{c}{r^{3\epsilon+1}}
+\frac{8\pi P r^{2}}{3}.
\label{eq:metric_pressure}
\end{equation}

The event horizon $r_h$ is determined by the largest positive root of
\begin{equation}
f(r_h)=0.
\label{eq:horizon_condition}
\end{equation}
Solving Eq.~\eqref{eq:horizon_condition} for the mass yields
\begin{equation}
M(r_h,P,Q)
=
\frac{r_h}{2}
\left[
1
+\frac{\omega^{2}Q^{2}}{r_h^{2}}
-\frac{c}{r_h^{3\epsilon+1}}
+\frac{8\pi P r_h^{2}}{3}
\right].
\label{eq:mass}
\end{equation}

In the extended phase space, the black hole mass is identified with the enthalpy,
\begin{equation}
H\equiv M.
\label{eq:enthalpy}
\end{equation}

\begin{figure}[htbp]
	\centering
	
	\includegraphics[width=0.48\textwidth]{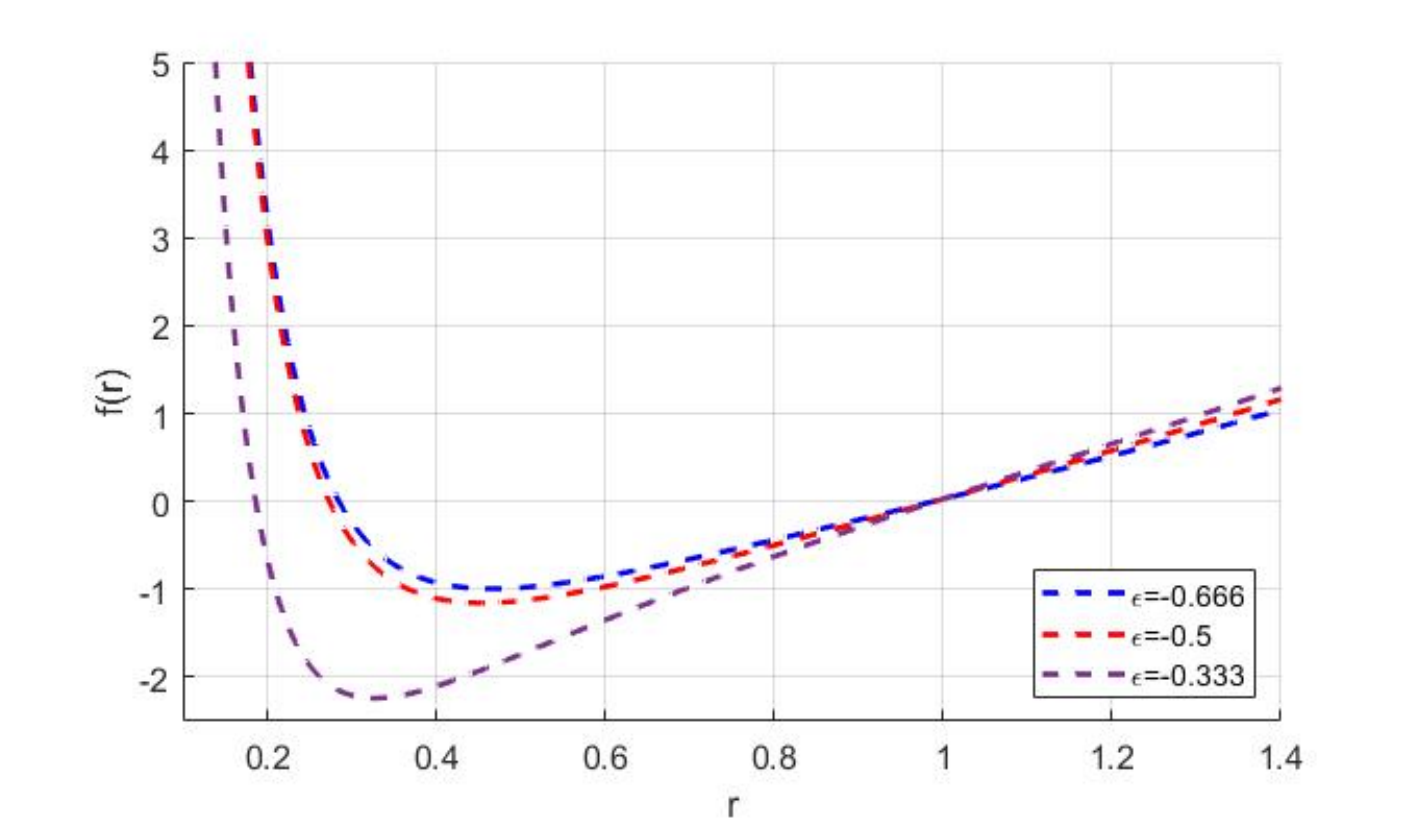}
	\hfill
	\includegraphics[width=0.48\textwidth]{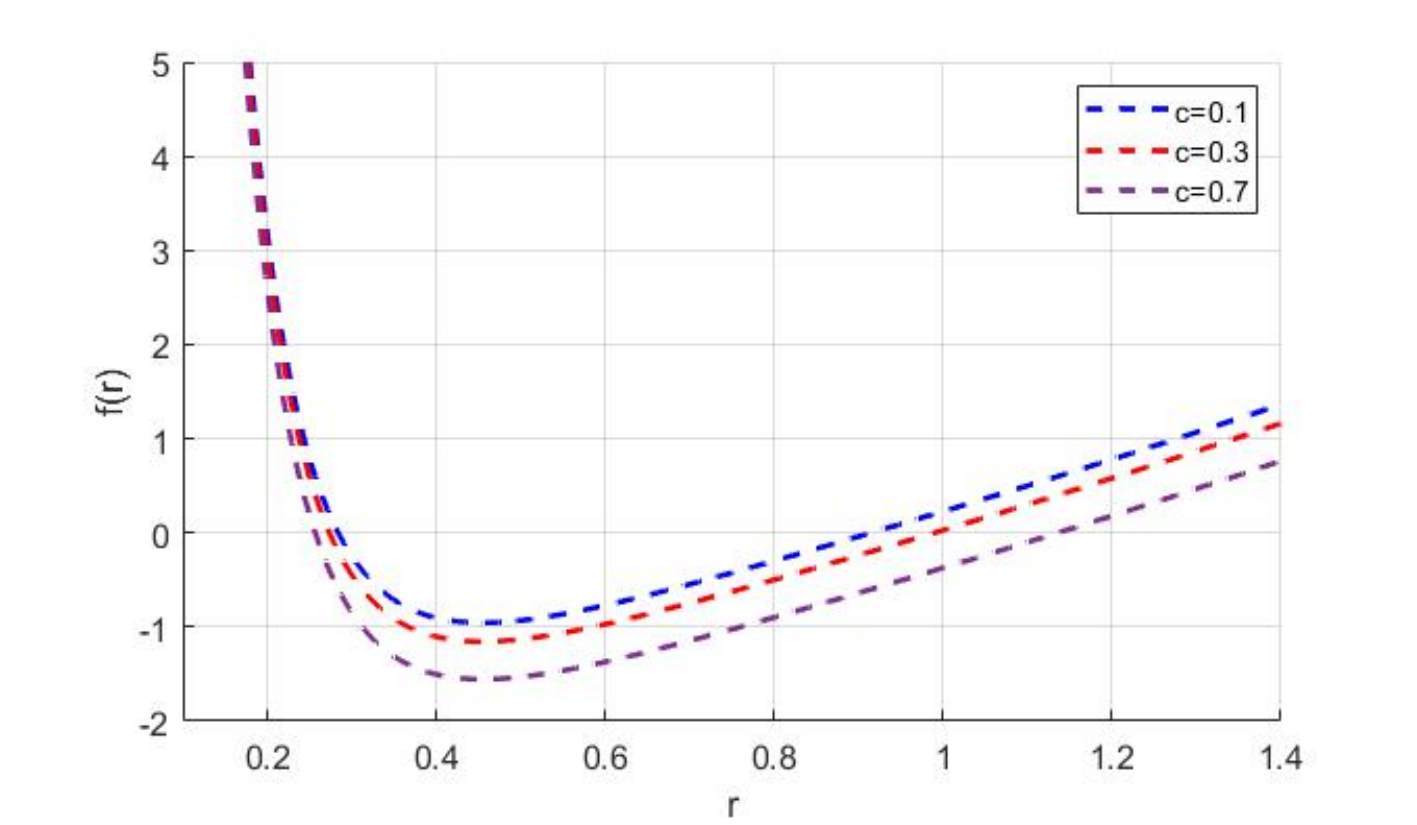}
	
	\vspace{0.3cm}
	
	\includegraphics[width=0.48\textwidth]{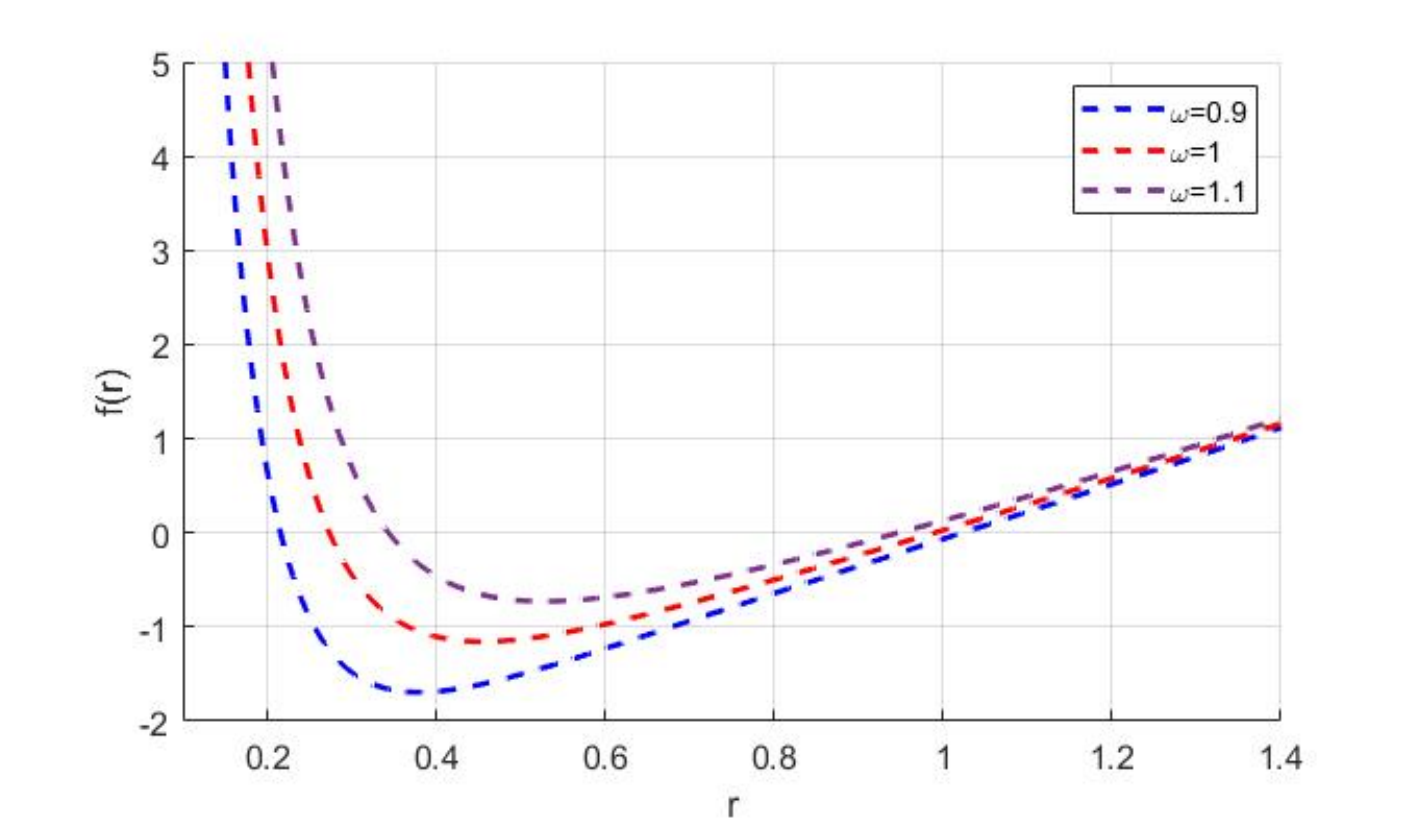}
	\hfill
	\includegraphics[width=0.48\textwidth]{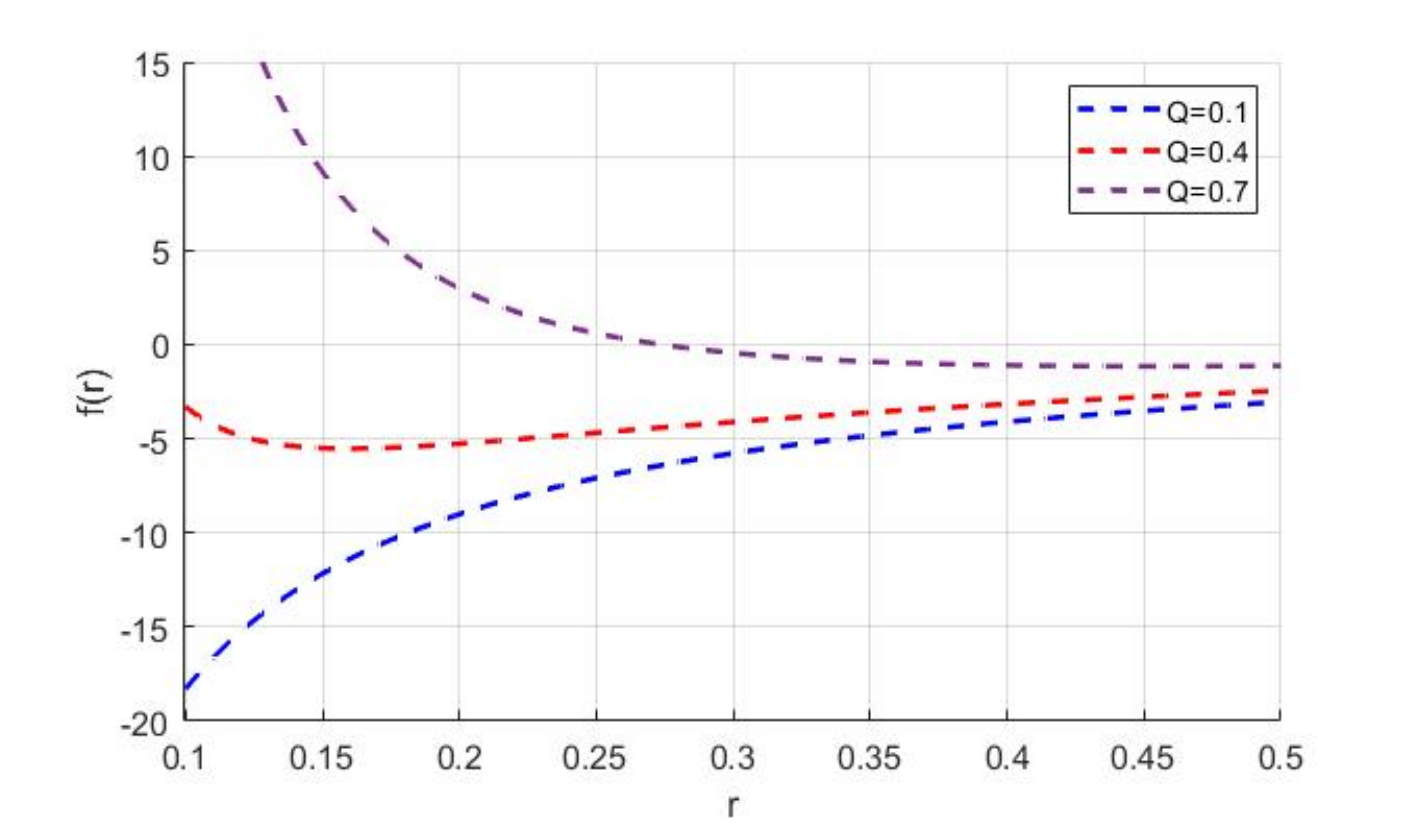}
	
	\vspace{0.3cm}
	
	\includegraphics[width=0.48\textwidth]{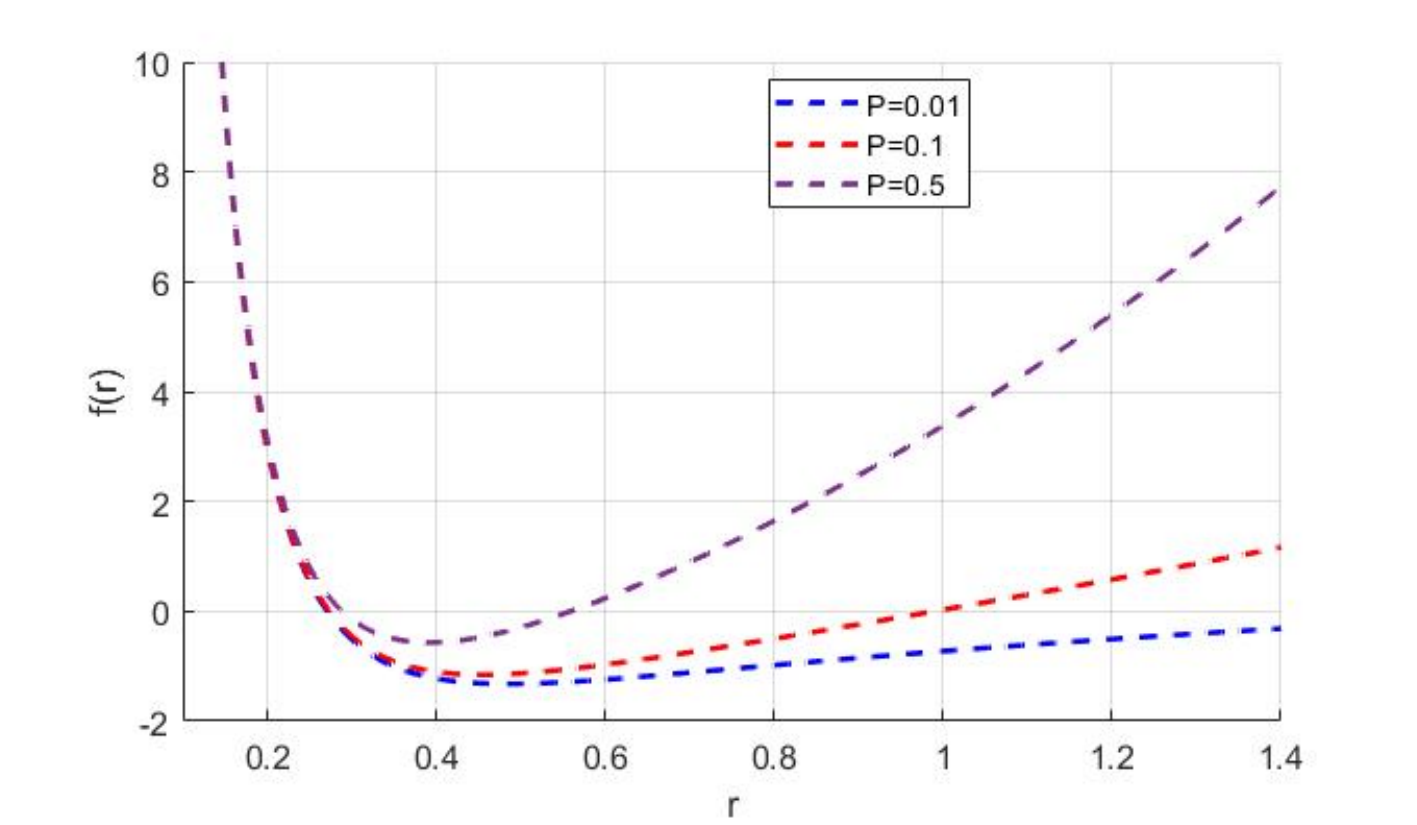}
	
	\caption{Variation of the metric function $f(r)$ with the horizon radius for different values of the quintessence parameter $\epsilon$, the quintessence normalization $c$, the effective NED parameter $\omega$, the dyonic charge $Q$, and the thermodynamic pressure $P$.}
	\label{F1}
\end{figure}
Fig.~1 illustrates the effects of $\epsilon$, $c$, $\omega$, $Q$, and $P$ on the metric function $f(r)$. The variations of these parameters modify the position and depth of the minimum of $f(r)$ and consequently affect the horizon structure. Less negative values of $\epsilon$ weaken the quintessence contribution, whereas increasing $c$ enhances its influence.
Increasing $\omega$ modifies the electromagnetic contribution and shifts the minimum of the metric function, while a larger dyonic charge increases the electromagnetic contribution. The pressure $P$ modifies the asymptotic AdS contribution and consequently affects the horizon structure. These results
show that the combined effects of nonlinear electrodynamics and quintessence can significantly deform the geometry of the charged AdS black hole.

\subsection{Thermodynaics}

The event horizon $r_h$ is determined by
\begin{equation}
f(r_h)=0.
\label{eq:horizon_condition}
\end{equation}
Solving Eq.~(\ref{eq:horizon_condition}) for the mass gives
\begin{equation}
M(r_h)rac{r_h}{2}
\left[
1+\frac{\omega^2Q^2}{r_h^2}
-\frac{c}{r_h^{3\epsilon+1}}
+\frac{8\pi P r_h^2}{3}
\right].
\label{eq:mass_rh}
\end{equation}

In the extended phase space formalism, $M$ is interpreted as the
thermodynamic enthalpy. 

The Bekenstein-Hawking entropy is given by one quarter of the horizon area [4,5],
\begin{equation}
S=\frac{A}{4},
\qquad
A=4\pi r_h^2,
\end{equation}
and hence
\begin{equation}
S=\pi r_h^2.
\label{eq:entropy}
\end{equation}
Consequently,
\begin{equation}
r_h=\sqrt{\frac{S}{\pi}}.
\label{eq:rh_entropy}
\end{equation}

The Hawking temperature is obtained from the surface gravity,
\begin{equation}
T
=
\frac{f'(r_h)}{4\pi}.
\label{eq:Hawking_temperature_definition}
\end{equation}
Using Eq.~\eqref{eq:metric_pressure}, we obtain
\begin{equation}
T
=
\frac{1}{4\pi}
\left[
\frac{1}{r_h}
-\frac{\omega^{2}Q^{2}}{r_h^{3}}
+8\pi P r_h
+3\epsilon c\,r_h^{-3\epsilon-2}
\right].
\label{eq:Hawking_temperature}
\end{equation}

The same temperature follows from the enthalpy through [4-9]
\begin{equation}
T=
\left(\frac{\partial M}{\partial S}\right)_{P,Q},
\end{equation}
which confirms the thermodynamic consistency of the solution.

In terms of the entropy, Eq.~(\ref{eq:Hawking_temperature}) becomes
\begin{equation}
T(S)
=
\frac{1}{4\sqrt{\pi S}}
\left[
1-\frac{\omega^2\pi Q^2}{S}
+3\epsilon c
\left(\frac{\pi}{S}\right)^{\frac{3\epsilon+1}{2}}
+8PS
\right].
\label{eq:temperature_entropy}
\end{equation}

\begin{figure}[htbp]
	\centering
	
	\includegraphics[width=0.48\textwidth]{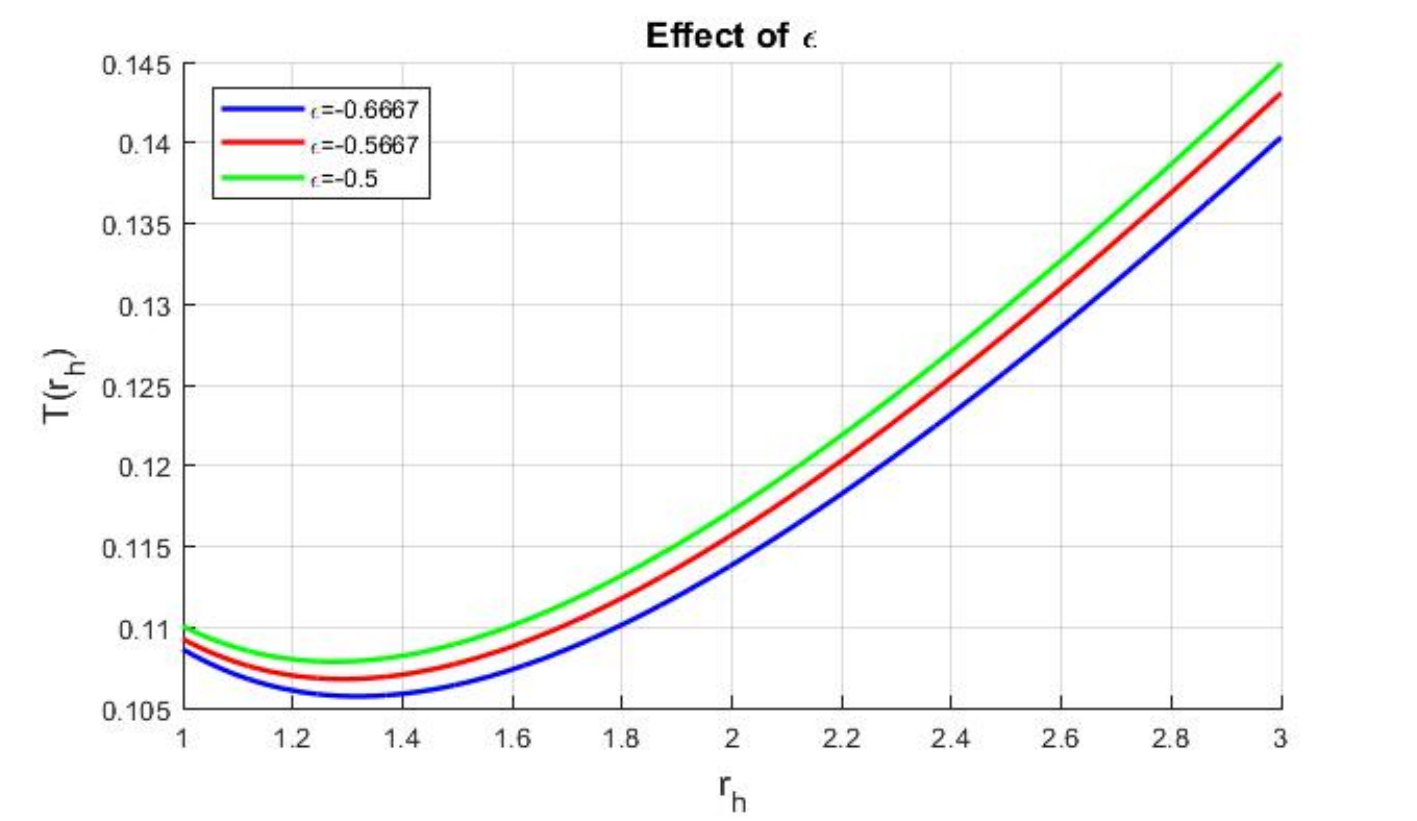}
	\hfill
	\includegraphics[width=0.48\textwidth]{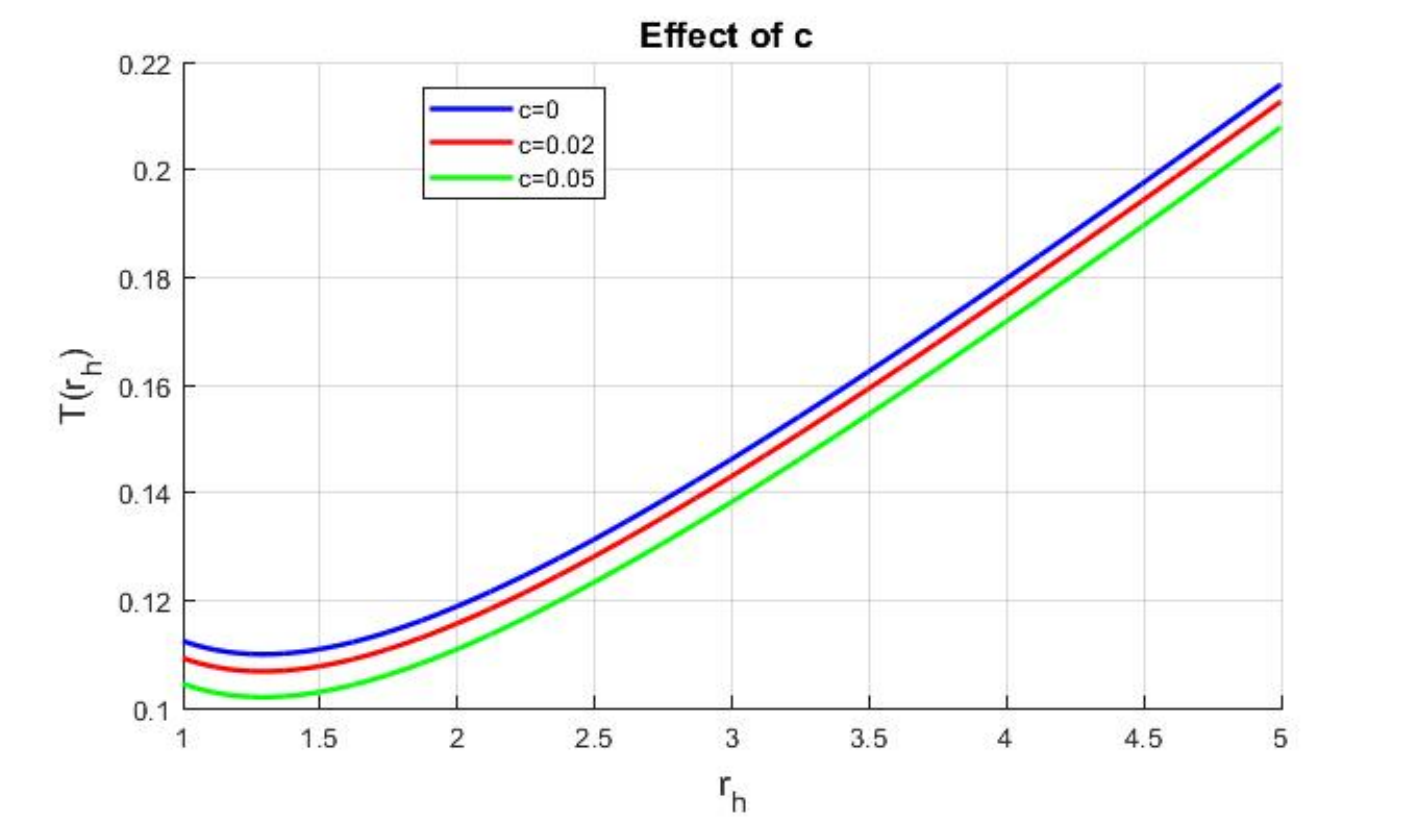}
	
	\vspace{0.3cm}
	
	\includegraphics[width=0.48\textwidth]{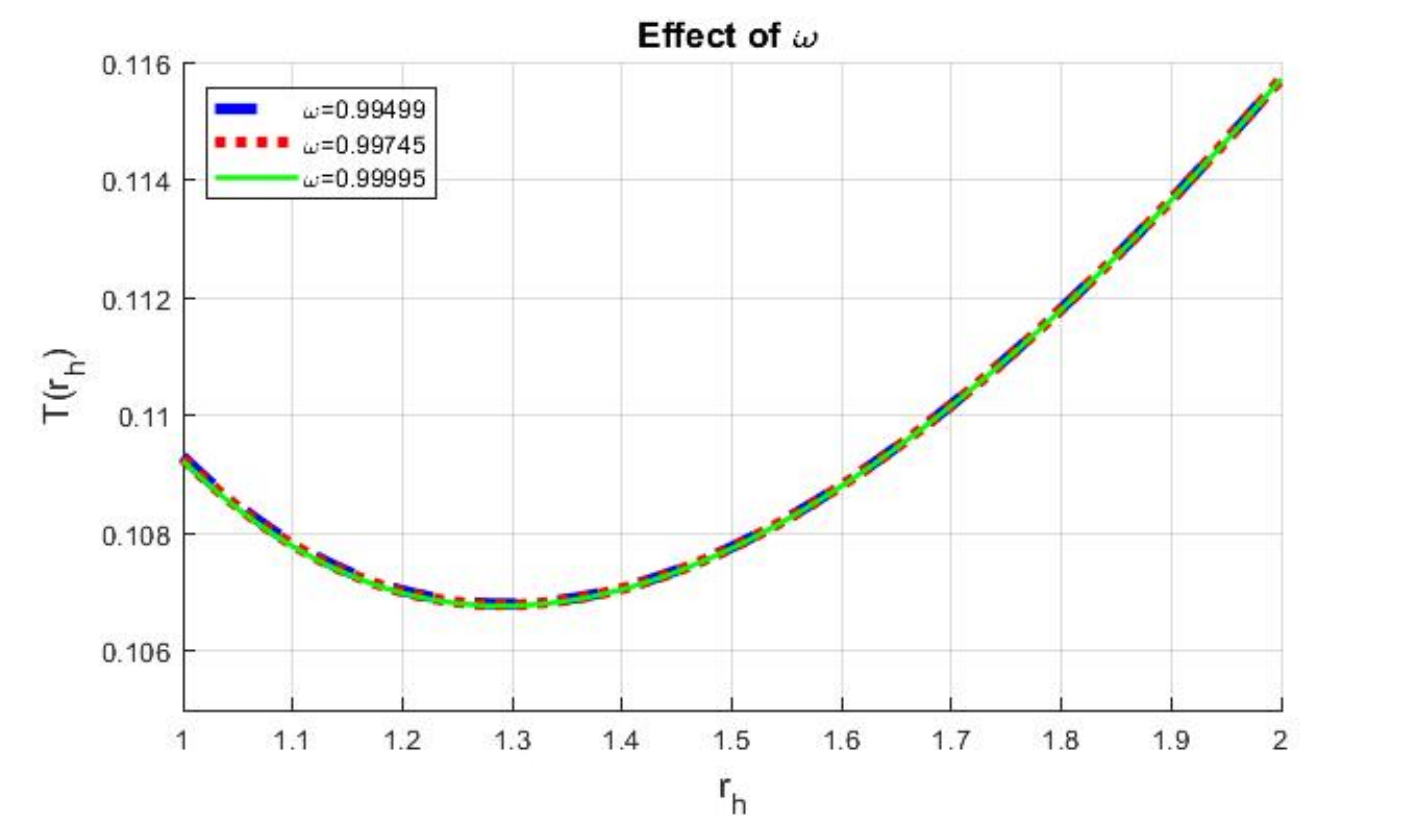}
	\hfill
	\includegraphics[width=0.48\textwidth]{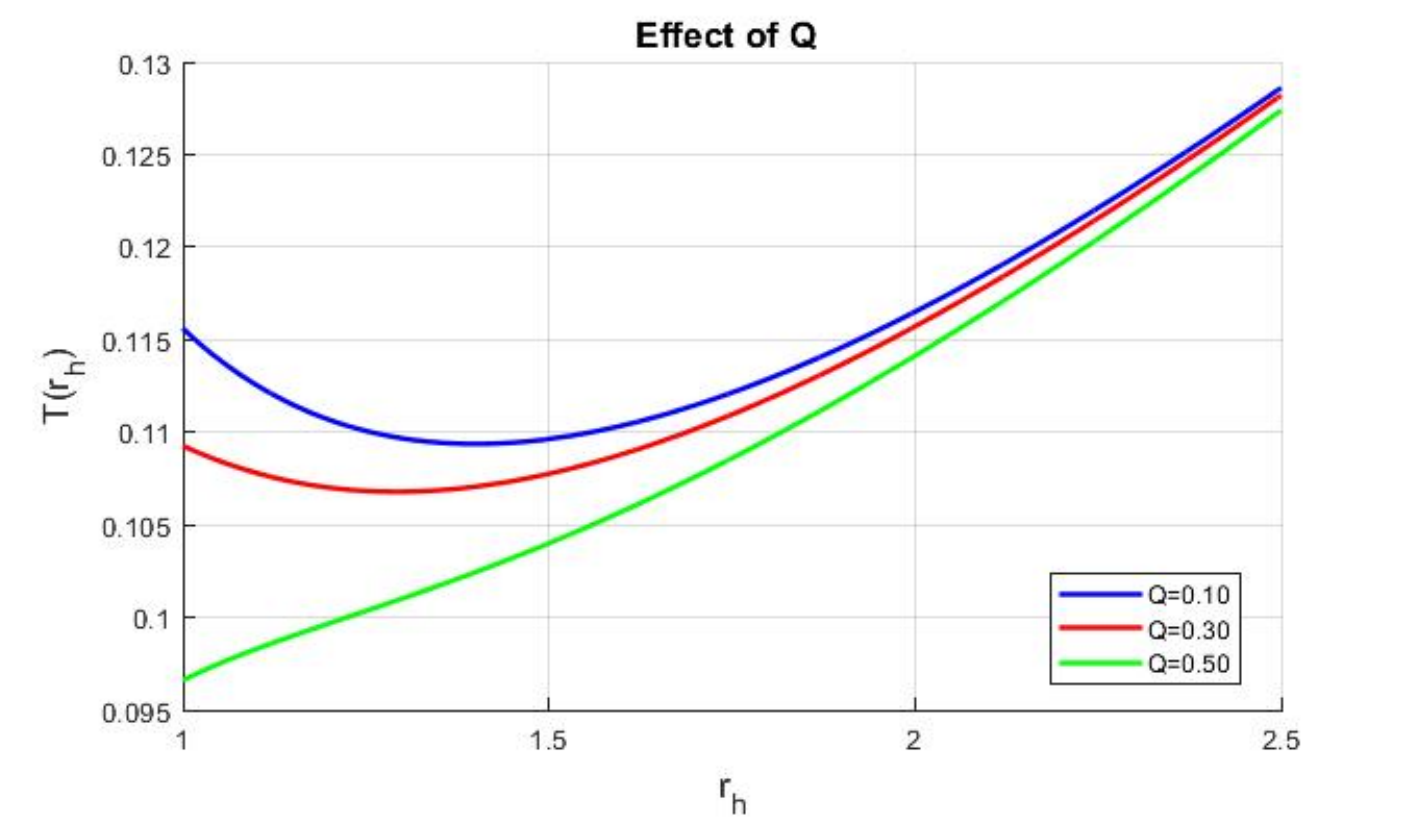}
	
	\caption{Hawking temperature as a function of the horizon radius for different values of the model parameters $\epsilon$, $c$, $\omega$, and $Q$.}
	\label{F2}
\end{figure}

The temperature profiles shown in Fig.~2 exhibit the characteristic extrema associated with the phase structure of charged AdS black holes. The quintessence contribution generally lowers the temperature and shifts the extrema, while increasing $c$ enhances this cooling effect. In contrast, variations of $\omega$ modify the electromagnetic contribution and shift the temperature extrema. Increasing the dyonic charge $Q$ strengthens the electromagnetic contribution and modifies the unstable branch. These behaviours continuously approach the corresponding Maxwell and no-quintessence limits when $\omega\sim 1$ and $c\sim 0$, respectively.

The thermodynamic volume associated with the black hole is defined by [4-9]
\begin{equation}
V= \left(\frac{\partial M}{\partial P}\right)_{S,Q}.
\label{eq:volume_definition}
\end{equation}
Using Eq.~(\ref{eq:mass}), one obtains
\begin{equation}
V
=
\frac{4}{3}\pi r_h^3,
\label{eq:volume_rh}
\end{equation}
or, in terms of the entropy,
\begin{equation}
V(S)
=
\frac{4}{3}\pi
\left(\frac{S}{\pi}\right)^{3/2}.
\label{eq:volume_entropy}
\end{equation}

The equation of state follows directly from Eq.~(\ref{eq:temperature_rh}) by solving for the pressure. We obtain
\begin{equation}
P
=
\frac{T}{2r_h}
-\frac{1}{8\pi r_h^2}
+\frac{\omega^2Q^2}{8\pi r_h^4}
-\frac{3\epsilon c}{8\pi r_h^{3\epsilon+3}}.
\label{eq:eos_rh}
\end{equation}
This relation explicitly describes the dependence of the pressure on the temperature and horizon radius. Using
\begin{equation}
r_h=\left(\frac{3V}{4\pi}\right)^{1/3},
\end{equation}
the equation of state can alternatively be expressed as
\begin{equation}
P(V,T)
=
\frac{T}{2}
\left(\frac{4\pi}{3V}\right)^{1/3}
-\frac{1}{8\pi}
\left(\frac{4\pi}{3V}\right)^{2/3}
+\frac{\omega^2Q^2}{8\pi}
\left(\frac{4\pi}{3V}\right)^{4/3}
-\frac{3\epsilon c}{8\pi}
\left(\frac{4\pi}{3V}\right)^{\epsilon+1}.
\label{eq:eos_VT}
\end{equation}

\begin{figure}[htbp]
	\centering
	\includegraphics[width=0.48\columnwidth]{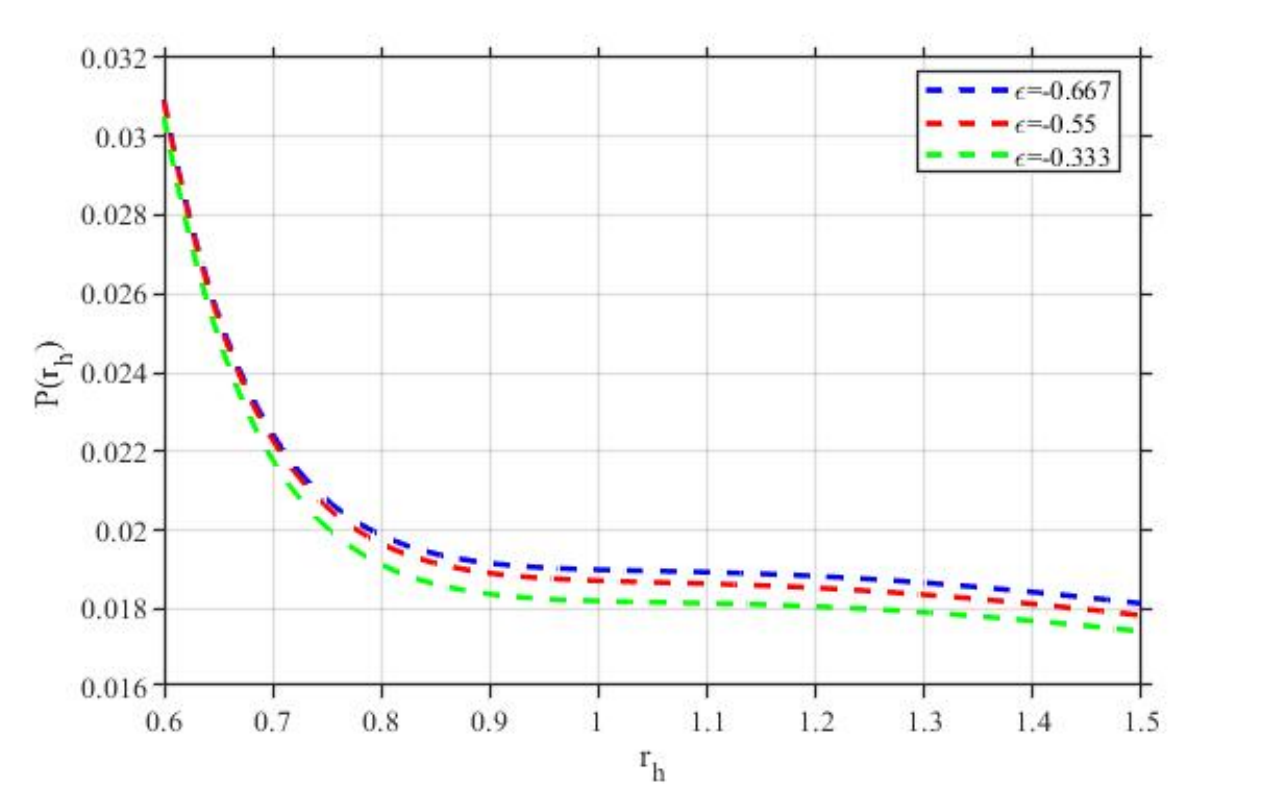}
	\hfill
	\includegraphics[width=0.48\columnwidth]{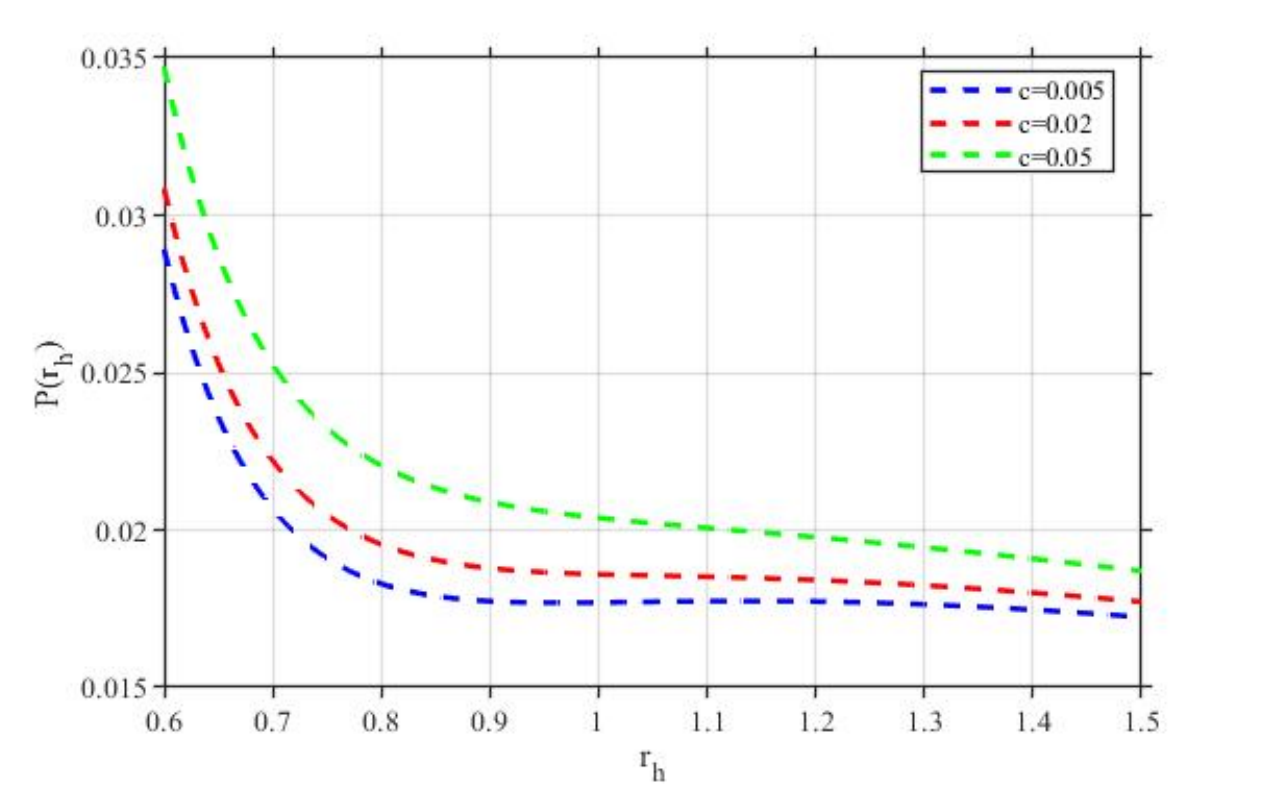}
	
	\vspace{0.2cm}
	
	\includegraphics[width=0.48\columnwidth]{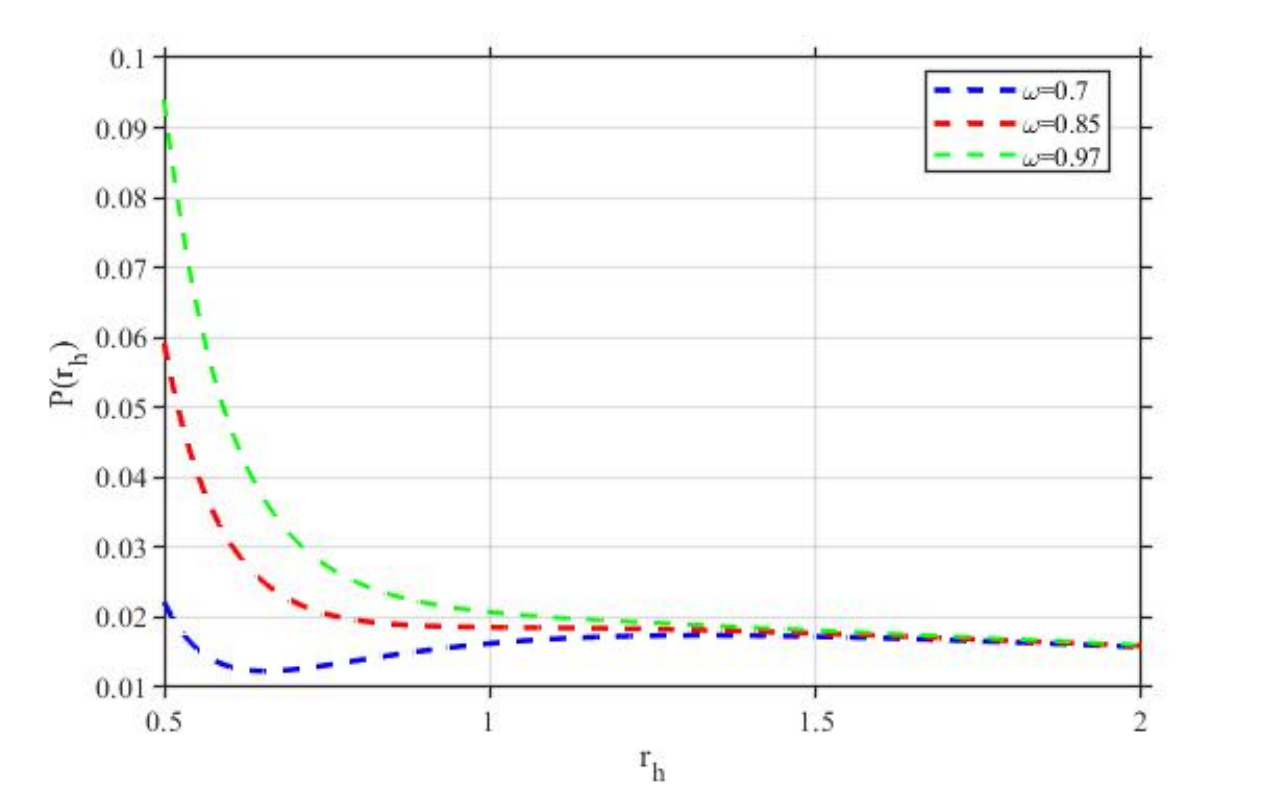}
	\hfill
	\includegraphics[width=0.48\columnwidth]{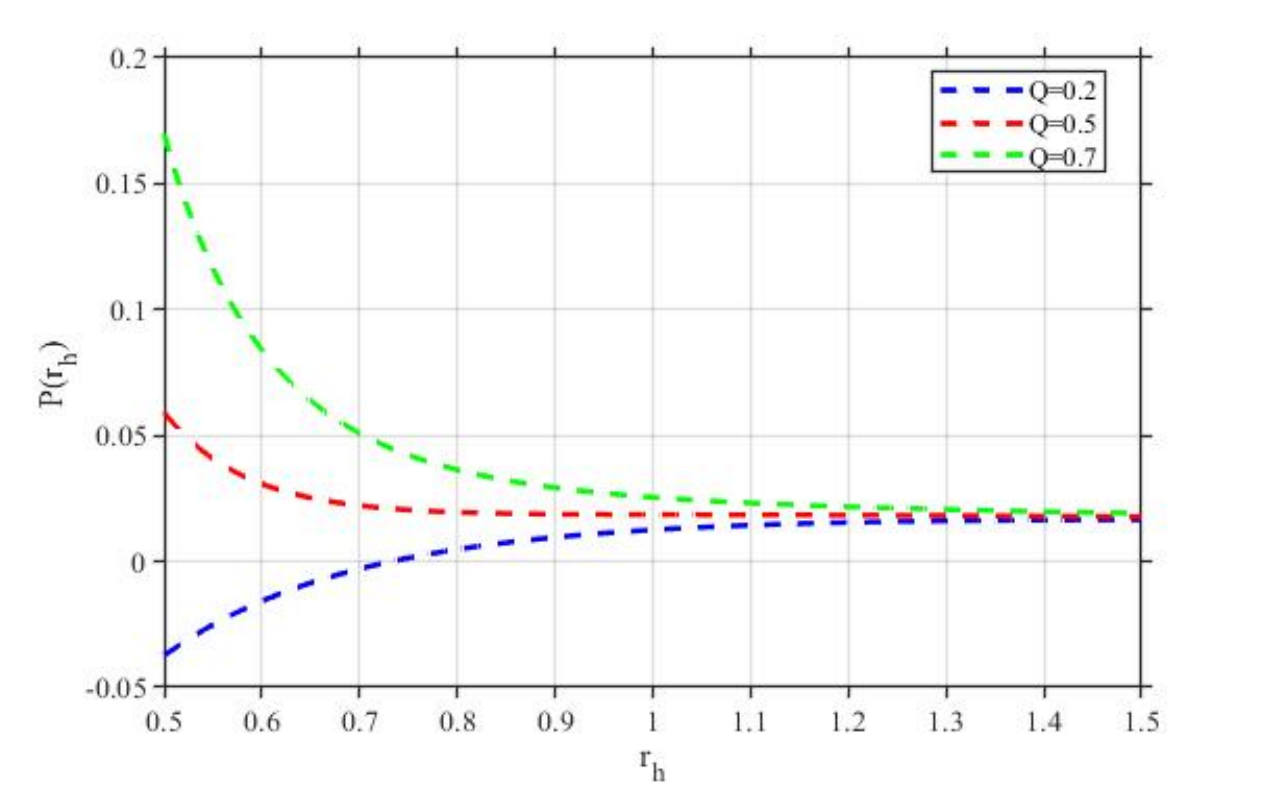}
	
	\caption{Pressure profile as a function of the horizon radius for different values of the quintessence parameter $\epsilon$, the quintessence normalization $c$, the effective NED parameter $\omega$ and the dyonic charge $Q$.}
	\label{F3}
\end{figure}

Fig.~3 presents the pressure profiles $P(r_h)$ for different values of $\epsilon$, $c$, $\omega$, and $Q$. The oscillatory behaviour of the isotherms is characteristic of a Van der Waals-like phase structure.
Increasing the quintessence contribution modifies the amplitude and position of the oscillatory region, whereas variations of $\omega$ shift the critical structure through the nonlinear electromagnetic contribution. Increasing the dyonic charge enhances the electromagnetic term and strengthens the oscillatory behaviour. In the limit $\omega\sim 1$, the corresponding linear-electrodynamics behaviour is recovered.

The local thermodynamic stability is characterized by the heat capacity at constant pressure [6-9],
\begin{equation}
C_P
=
T\left(\frac{\partial S}{\partial T}\right)_P.
\label{eq:cp_definition}
\end{equation}
Using Eqs.~(\ref{eq:entropy}) and
(\ref{eq:temperature_entropy}), we obtain
\begin{equation}
C_P
=
\frac{
	2\pi r_h^2
	\left[
	r_h\left(
	r_h+8\pi P r_h^3
	+3\epsilon c\,r_h^{-3\epsilon}
	\right)
	-\omega^2Q^2
	\right]
}{
-r_h+8\pi P r_h^3
-3\epsilon(3\epsilon+2)c\,r_h^{-3\epsilon+1}
+3\omega^2Q^2/r_h
}.
\label{eq:heat_capacity}
\end{equation}

\begin{figure}[htbp]
	\centering
	\includegraphics[width=0.48\textwidth]{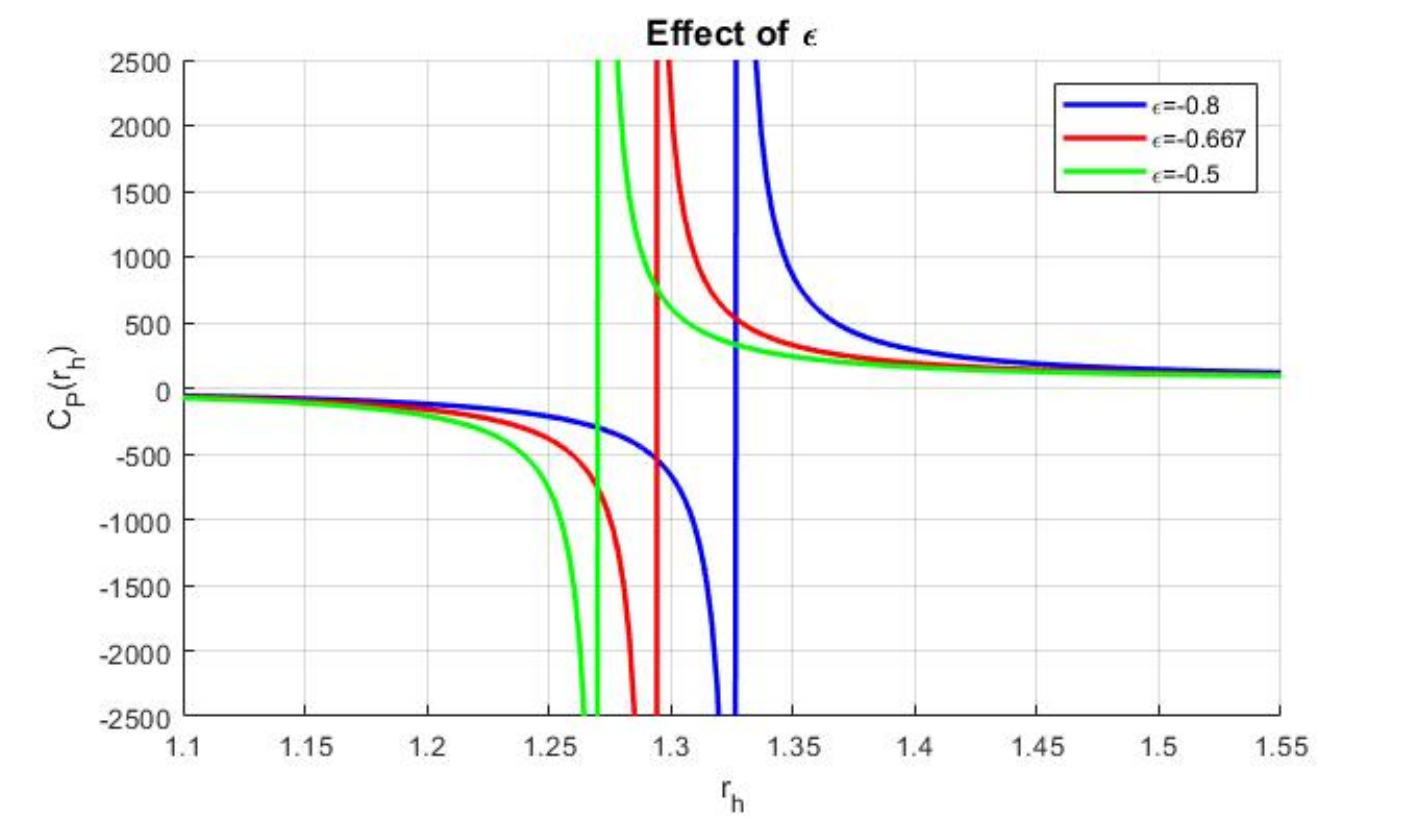}
	\hfill
	\includegraphics[width=0.48\textwidth]{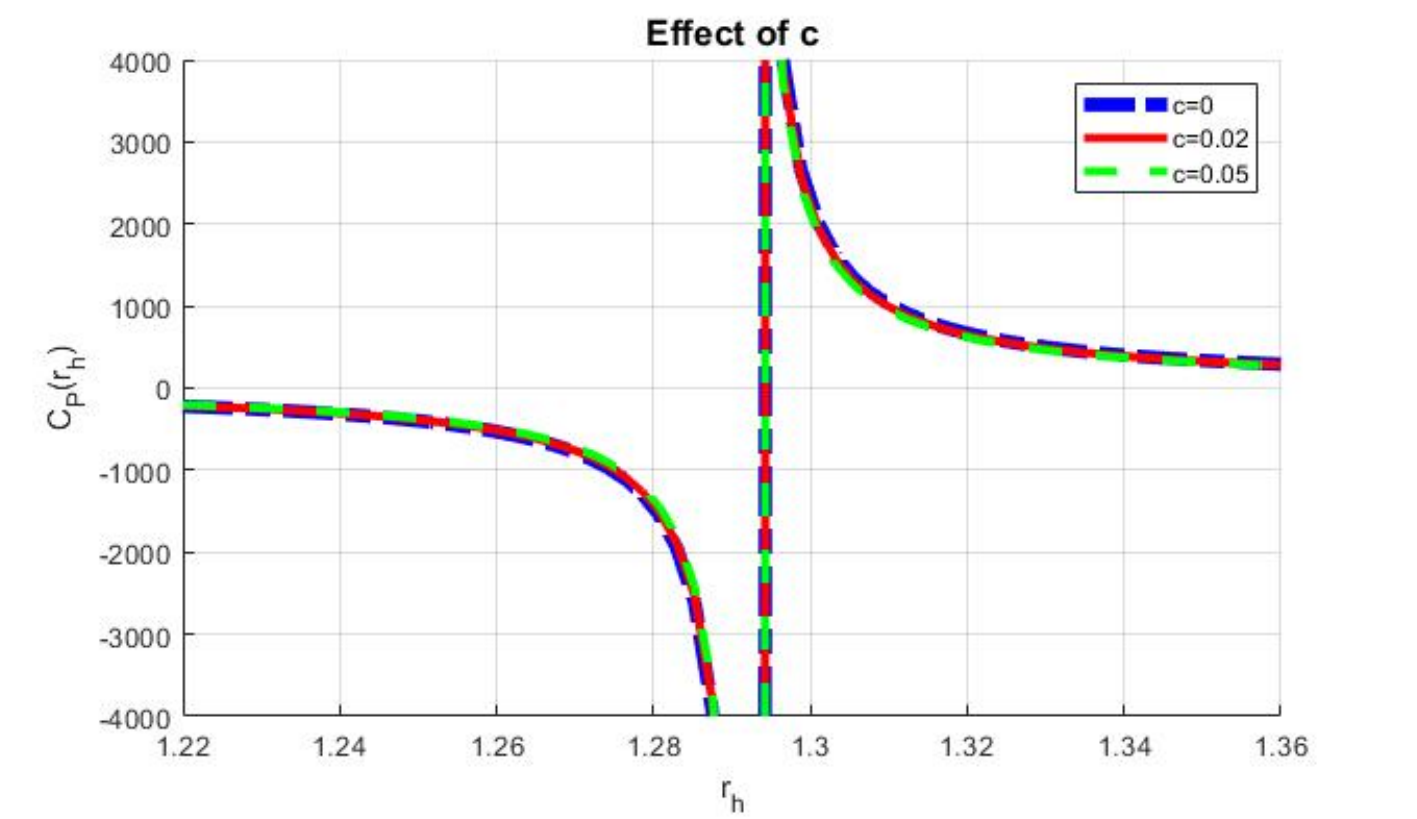}
	
	\vspace{0.3cm}
	
	\includegraphics[width=0.48\textwidth]{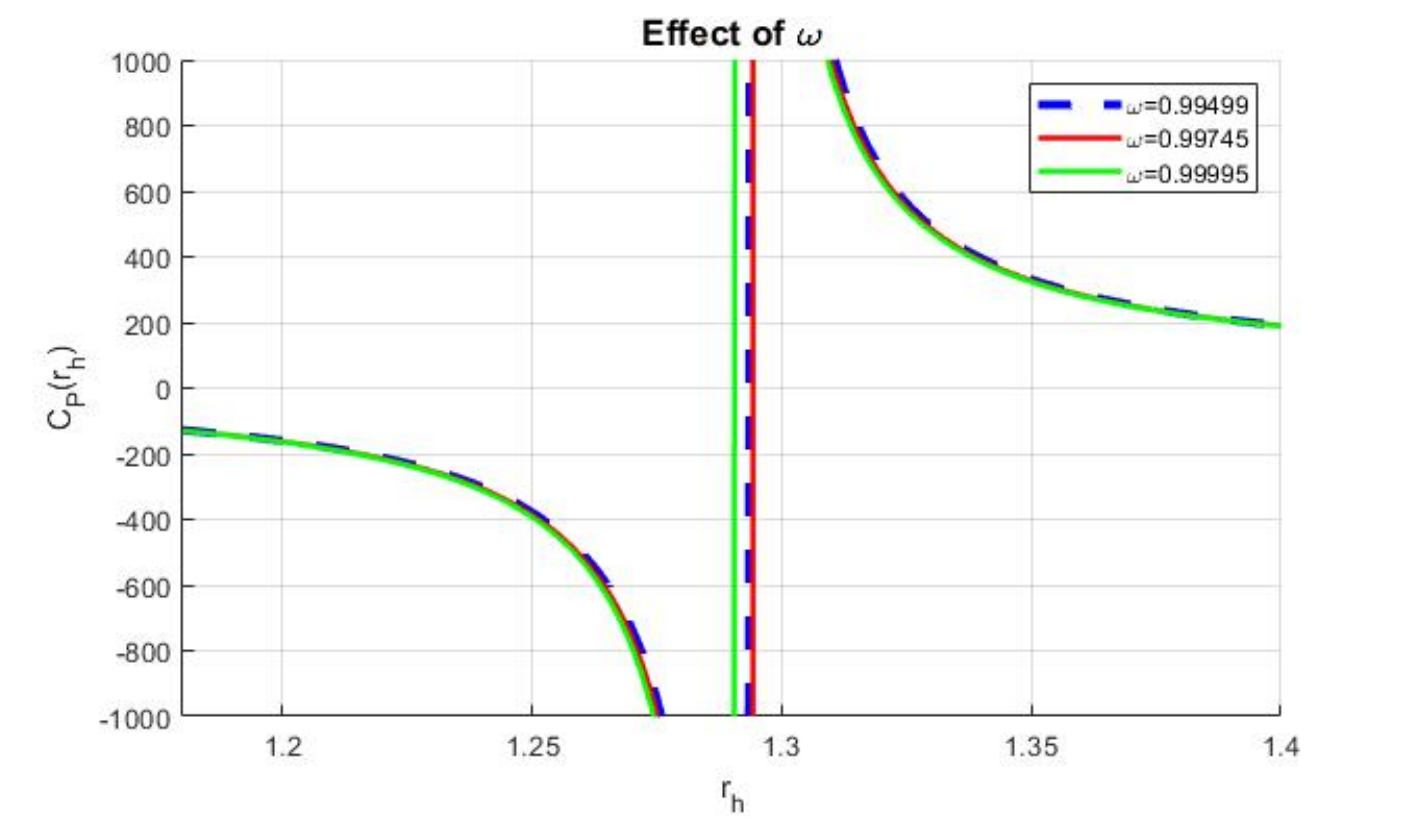}
	\hfill
	\includegraphics[width=0.48\textwidth]{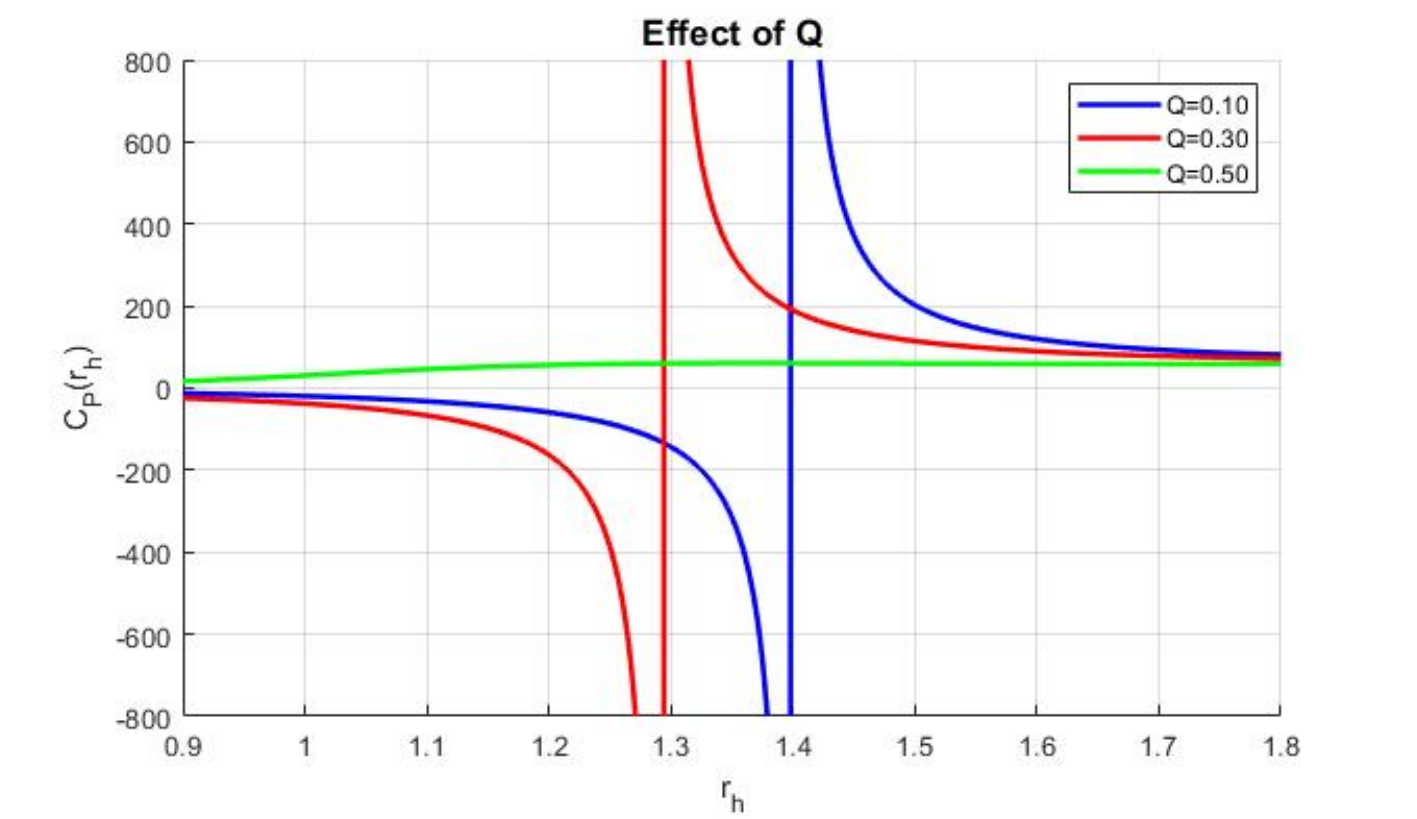}
	
	\vspace{0.3cm}
	
	\includegraphics[width=0.48\textwidth]{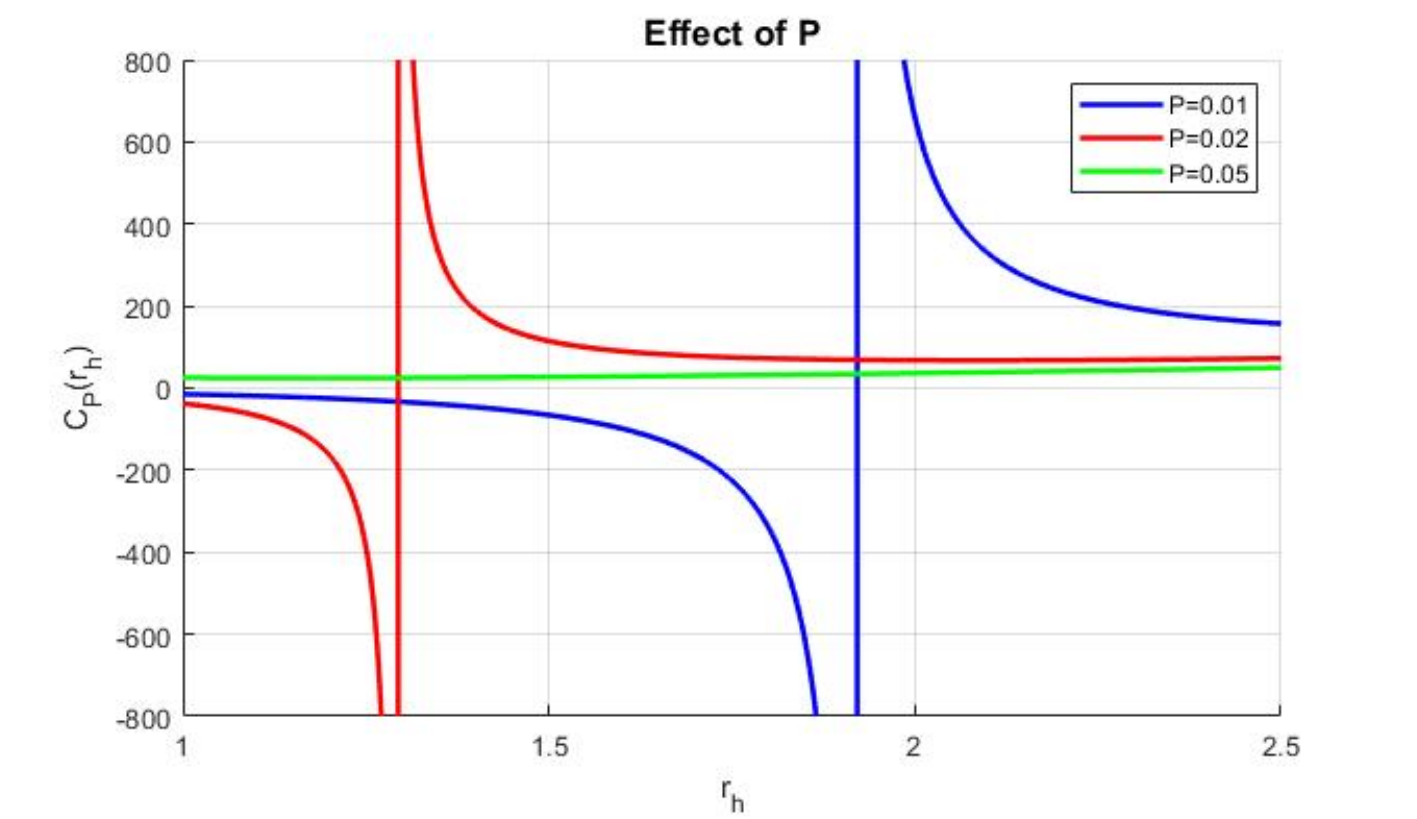}
	
	\caption{Behavior of the heat capacity as a function of the horizon	radius for different values of the quintessence parameter $\epsilon$, the quintessence normalization $c$, the effective NED parameter $\omega$, the dyonic charge $Q$, and the thermodynamic pressure $P$.}
	\label{F4}
\end{figure}

The divergences of $C_P$ signal the boundaries between thermodynamically stable and unstable branches and are associated with second-order critical behaviour. Fig.~4 shows the dependence of $C_P$ on the horizon radius. The positions and amplitudes of the divergences are modified by all the
model parameters. Variations of $\epsilon$ and $c$ shift the instability regions, while $\omega$ changes the electromagnetic contribution. Increasing the dyonic charge enhances the electromagnetic contribution and modifies the strength of the heat capacity divergences. The pressure also shifts the
location of the critical peaks.

The global thermodynamic behaviour is further characterized by the Gibbs free energy,
\begin{equation}
G=M-TS,
\end{equation}
which gives
\begin{equation}
G
=
\frac{r_h}{4}
\left[
1+\frac{3\omega^2Q^2}{r_h^2}
-\frac{(3\epsilon+1)c}{r_h^{3\epsilon+1}}
-\frac{8\pi P r_h^2}{3}
\right].
\label{eq:gibbs}
\end{equation}

The Gibbs free energy is particularly useful for analysing global phase behaviour. Multiple branches and swallow-tail-type structures, when present, are associated with first-order phase transitions.

\begin{figure}[htbp]
	\centering
	\includegraphics[width=0.48\textwidth]{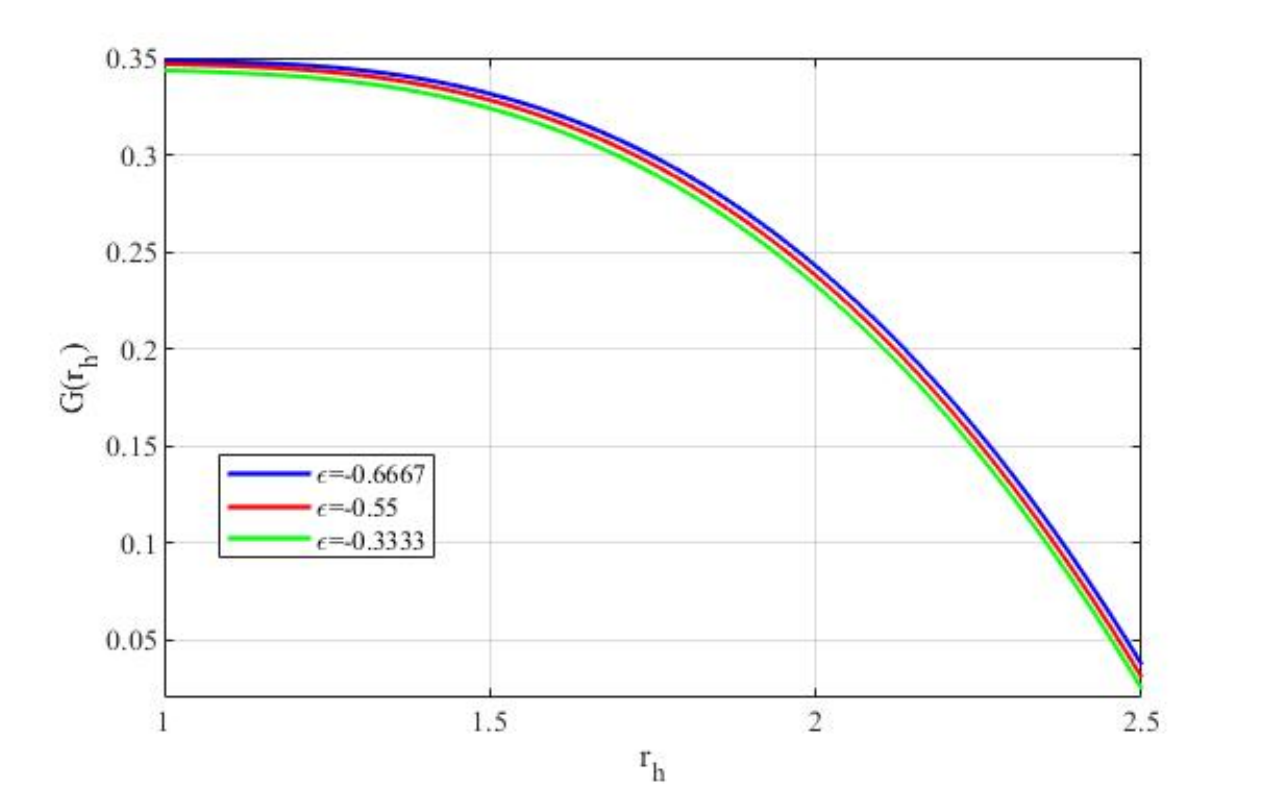}
	\hfill
	\includegraphics[width=0.48\textwidth]{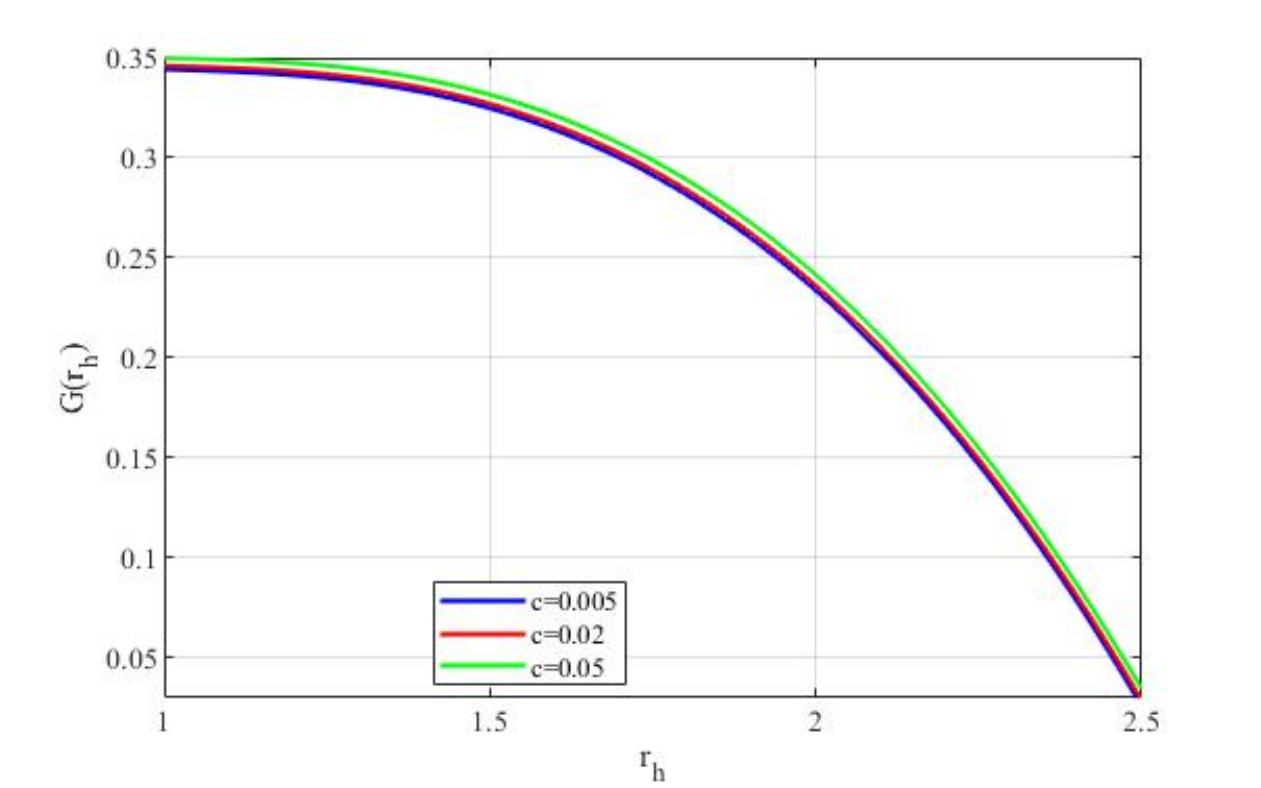}
	
	\vspace{0.3cm}
	
	\includegraphics[width=0.48\textwidth]{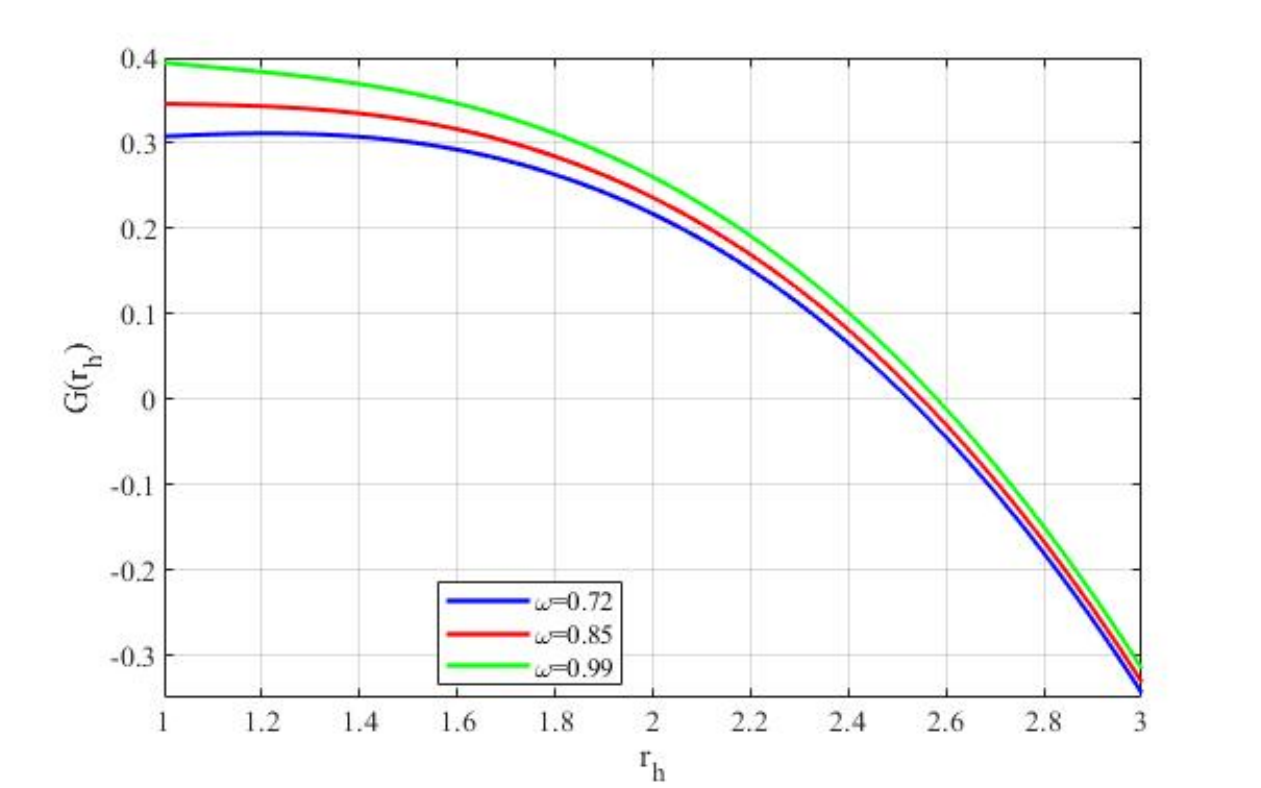}
	\hfill
	\includegraphics[width=0.48\textwidth]{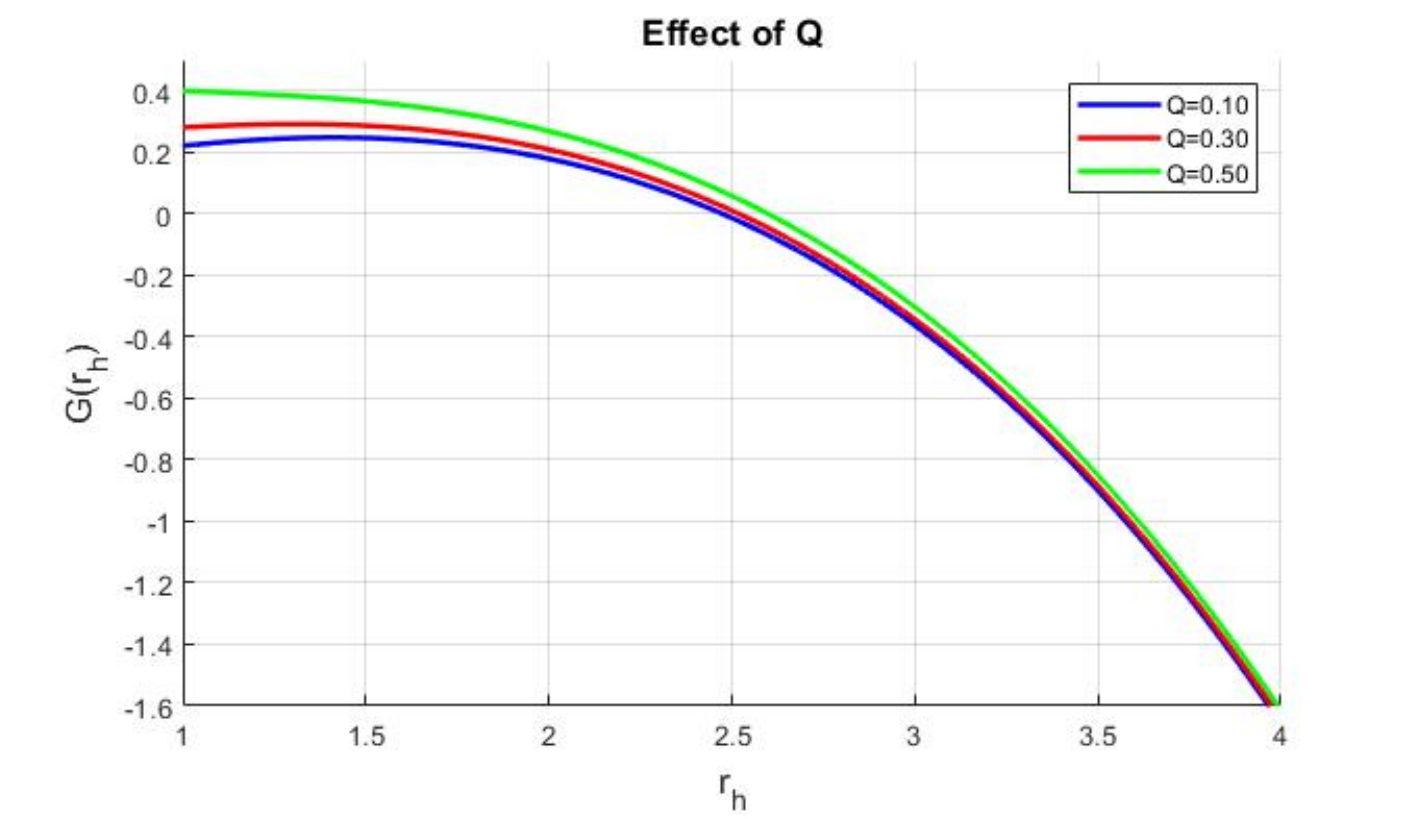}
	
	\vspace{0.3cm}
	
	\includegraphics[width=0.48\textwidth]{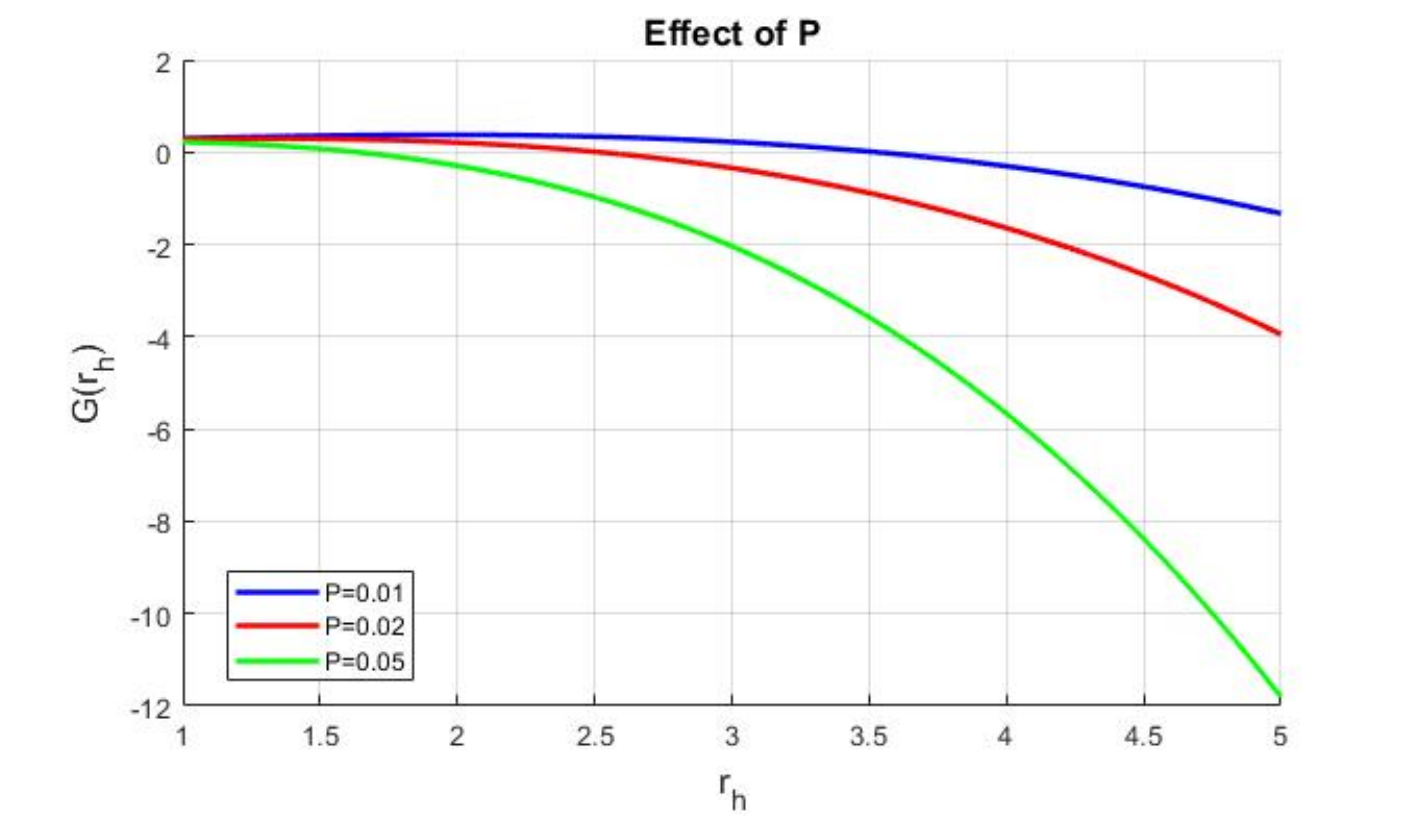}
	
	\caption{Gibbs free energy behavior versus of the horizon radius for different values of the quintessence parameter $\epsilon$, the quintessence normalization $c$, the effective NED parameter $\omega$, the dyonic charge $Q$, and the thermodynamic pressure $P$.}
	\label{F5}
\end{figure}
	
Fig.~5 displays $G(r_h)$ for different values of $\epsilon$, $c$, $\omega$, $Q$, and $P$.
The quintessence parameters modify the free-energy landscape and shift the relative position of the thermodynamic branches. Variations of $\omega$ and $Q$ alter the electromagnetic contribution and consequently modify the multibranch structure, while increasing $P$ changes the separation and position of the branches. The Gibbs free energy therefore complements the
heat capacity analysis by providing information about the global
thermodynamic behaviour.

The critical points are determined by the inflection conditions
\begin{equation}
\left(\frac{\partial P}{\partial r_h}\right)_T=0,
\qquad
\left(\frac{\partial^2P}{\partial r_h^2}\right)_T=0.
\label{eq:critical_conditions}
\end{equation}
Using the equation of state (26), these conditions yield an implicit equation for the critical horizon radius:
\begin{equation}
2r_c^2
+
3\epsilon c(3\epsilon+3)(3\epsilon+2)
r_c^{\,1-3\epsilon}
-
12\omega^2Q^2
=0.
\label{eq:critical_radius}
\end{equation}
The corresponding critical temperature is
\begin{equation}
T_c
=
\frac{1}{2\pi r_c}
-\frac{\omega^2Q^2}{\pi r_c^3}
+\frac{3\epsilon(3\epsilon+3)c}
{4\pi}
r_c^{-(3\epsilon+2)},
\label{eq:critical_temperature}
\end{equation}
while the critical pressure is
\begin{equation}
P_c
=
\frac{1}{8\pi r_c^2}
-\frac{\omega^2Q^2}{8\pi r_c^4}
+\frac{3\epsilon c(3\epsilon+2)}
{8\pi r_c^{3\epsilon+3}}.
\label{eq:critical_pressure}
\end{equation}

In the absence of quintessence, $c=0$, Eq.~(\ref{eq:critical_radius})
reduces to
\begin{equation}
r_c^2=6\omega^2Q^2,
\end{equation}
and consequently
\begin{equation}
T_c
=
\frac{1}{3\sqrt{6}\pi\omega Q},
\qquad
P_c
=
\frac{1}{96\pi\omega^2Q^2}.
\label{eq:critical_limit}
\end{equation}

The Maxwell limit is obtained by taking $\gamma\sim 0$. In this limit, the NED Lagrangian reduces to the linear Maxwell form and Eq.~\eqref{eq:omega_definition} gives $\omega\sim 1$. Consequently, the electromagnetic contribution reduces to the standard dyonic Reissner-Nordström term $Q^2/r^2$.

In the thermodynamic analysis below, $\omega$ is treated as a fixed effective NED parameter, corresponding to fixed values of the underlying nonlinear coupling and charge ratio. This allows us to isolate the thermodynamic effects of the effective nonlinear electromagnetic correction.

The physical parameters and their corresponding constraints entering the black hole solution and the subsequent thermodynamic analysis are summarized in Table~\ref{tab:parameters}.

\begin{table}[ht]
	\centering
	\caption{Model parameters and their physical ranges used in the thermodynamic analysis.}
	\label{tab:parameters}
	\begin{tabular}{c c c}
		\hline
		\textbf{Parameter} & \textbf{Meaning} & \textbf{Range/condition} \\
		\hline
		$\gamma$ 
		& NED coupling parameter 
		& $0<\gamma<\tanh^{-1}(\sqrt{2}/2)$ \\
		
		$q=Q_e/Q_m$
		& Electric-to-magnetic charge ratio
		& $q\geq0$ \\
		
		$Q_e$
		& Electric charge
		& $Q_e\geq0$ \\
		
		$Q_m$
		& Magnetic charge
		& $Q_m\geq0$ \\
		
		$Q^2=Q_e^2+Q_m^2$
		& Total dyonic charge
		& $Q>0$ \\
		
		$\omega$
		& Effective NED parameter
		& $\displaystyle
		\omega^2=
		\cosh\gamma-
		\frac{1-q^2}{1+q^2}\sinh\gamma$ \\
		
		$c$
		& Quintessence normalization
		& $c\geq0$ \\
		
		$\epsilon$
		& Quintessence equation-of-state parameter
		& $-1<\epsilon<-\frac{1}{3}$ \\
		
		$P$
		& Thermodynamic pressure
		& $P>0$ \\
		
		$r_h$
		& Event-horizon radius
		& $r_h>0$ \\
		
		$\Lambda$
		& Cosmological constant
		& $\Lambda<0$ \\
		
		$M$
		& Black hole mass/enthalpy
		& $M>0$ \\
		\hline
	\end{tabular}
\end{table}

In particular, $\omega$ is not an independent parameter of the underlying NED theory, but an effective quantity determined by the nonlinear coupling $\gamma$ and the electric-to-magnetic charge ratio $q=Q_e/Q_m$ through
Eq.~\eqref{eq:omega_definition}. In the Maxwell limit
$\gamma\sim 0$, one has $\omega\sim 1$.

\section{Extended Phase Space and Joule-Thomson Expansion}
\label{sec:JT}

The Joule-Thomson (JT) expansion describes an isenthalpic process in which the enthalpy remains constant while the pressure and temperature of the system change. In the extended phase-space description of AdS black holes, the mass $M$ is identified as the enthalpy of the black hole [7,39].
The JT expansion therefore provides a useful framework for analysing the thermal response of black holes and identifying the regions in which the system undergoes cooling or heating [39-49].

The central quantity characterizing the JT process is the Joule-Thomson
coefficient,
\begin{equation}
\mu=
\left(\frac{\partial T}{\partial P}\right)_{M},
\label{eq:28}
\end{equation}
which measures the variation of the temperature with pressure along an isenthalpic trajectory. A positive value of $\mu$ corresponds to cooling, whereas $\mu<0$ corresponds to heating. The boundary between these two regimes is determined by the inversion condition $\mu=0$, which defines the inversion temperature and pressure [39-49].

For the black hole considered here, the enthalpy is given by Eq.~(12), which can be written in terms of the horizon radius as
\begin{equation}
M=
\frac{r_h}{2}
\left(
1+\frac{\omega^2Q^2}{r_h^2}
-\frac{c}{r_h^{3\epsilon+1}}
+\frac{8\pi P r_h^2}{3}
\right).
\label{eq:29}
\end{equation}
The Hawking temperature is consequently
\begin{equation}
T=
\frac{1}{4\pi r_h^3}
\left[
r_h
\left(
r_h+8\pi P r_h^3
+3\epsilon c r_h^{-3\epsilon}
\right)
-\omega^2Q^2
\right].
\label{eq:30}
\end{equation}

The JT coefficient can alternatively be expressed in terms of the heat capacity and thermodynamic volume as
\begin{equation}
\mu
=
\left(\frac{\partial T}{\partial P}\right)_{M}
=
\frac{1}{C_P}
\left[
T\left(\frac{\partial V}{\partial T}\right)_P-V
\right],
\label{eq:31}
\end{equation}
where
\begin{equation}
V=\frac{4}{3}\pi r_h^3.
\end{equation}
This relation makes explicit the connection between the JT expansion and the thermodynamic stability encoded in $C_P$.

For a direct calculation, we impose the isenthalpic condition
\begin{equation}
dM=0.
\end{equation}
The total differential of the mass is then
\begin{equation}
dM=
\left(\frac{\partial M}{\partial P}\right)_{r_h}dP
+
\left(\frac{\partial M}{\partial r_h}\right)_Pdr_h=0.
\label{eq:32}
\end{equation}
Consequently,
\begin{equation}
\left(\frac{dr_h}{dP}\right)_M
=
-
\frac{
	\left(\frac{\partial M}{\partial P}\right)_{r_h}
}{
\left(\frac{\partial M}{\partial r_h}\right)_P
}.
\label{eq:33}
\end{equation}

The total variation of the Hawking temperature along an isenthalpic trajectory is
\begin{equation}
\left(\frac{\partial T}{\partial P}\right)_M
=
\left(\frac{\partial T}{\partial P}\right)_{r_h}
+
\left(\frac{\partial T}{\partial r_h}\right)_P
\left(\frac{dr_h}{dP}\right)_M .
\label{eq:34}
\end{equation}
Using Eqs.~(29)-(33), the JT coefficient can be written explicitly as
\begin{equation}
\mu
=
2r_h
-
\frac{
	\displaystyle
	\frac{r_h^3}{3}
	\left[
	-\,\frac{1}{r_h^2}
	+\frac{3\omega^2Q^2}{r_h^4}
	+8\pi
	-\frac{3\epsilon(3\epsilon+2)c}
	{r_h^{3\epsilon+3}}
	\right]
}{
\displaystyle
\frac{1}{2}
-\frac{\omega^2Q^2}{2r_h^2}
+\frac{3\epsilon c}
{2r_h^{3\epsilon+1}}
+4\pi P r_h^2
}.
\end{equation}

\begin{equation}
\mu = \frac{2 r_{h}\left[r_{h}(4r_{h} + 16{\pi P r_{h}^{3}} + 3\epsilon (3\epsilon + 5) c r_{h}^{-3\epsilon}) - 6(\omega Q)^{2}\right]}{3\left[r_{h}(r_{h} + 8{\pi P r_{h}^{3}} + 3\epsilon c r_{h}^{-3\epsilon})-(\omega Q)^{2}\right]}.
\label{eq:35}
\end{equation}

Equation~(\ref{eq:35}) explicitly shows that the JT response results from the competition between the electromagnetic, nonlinear-electrodynamics, quintessence, and AdS pressure contributions. It also makes clear that singular behaviour of the JT coefficient can be associated with the thermodynamic stability structure through the denominator, which is related
to the derivative of the enthalpy with respect to the horizon radius.

\begin{figure}[htbp]
	\centering
	
	\includegraphics[width=0.48\textwidth]{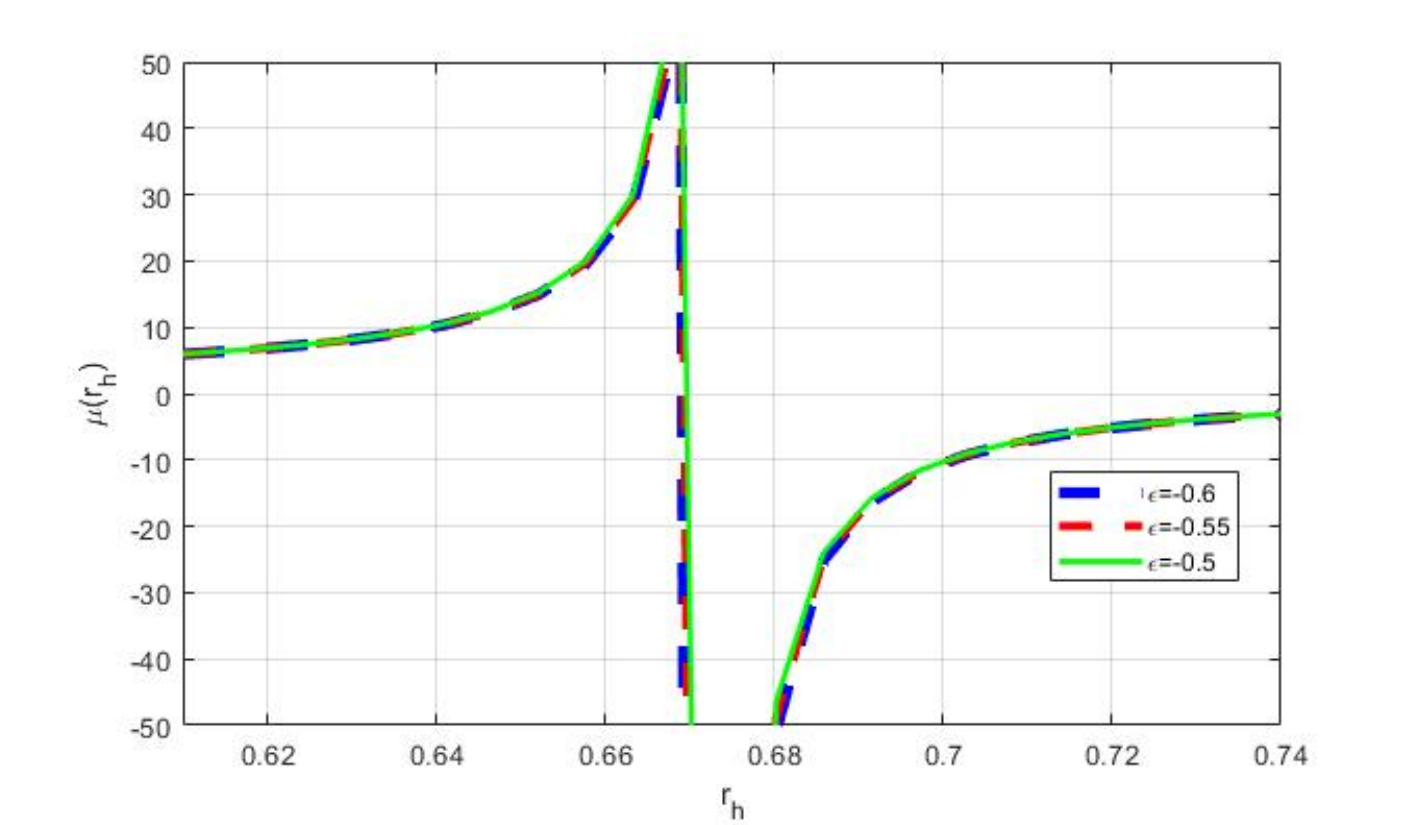}
	\hfill
	\includegraphics[width=0.48\textwidth]{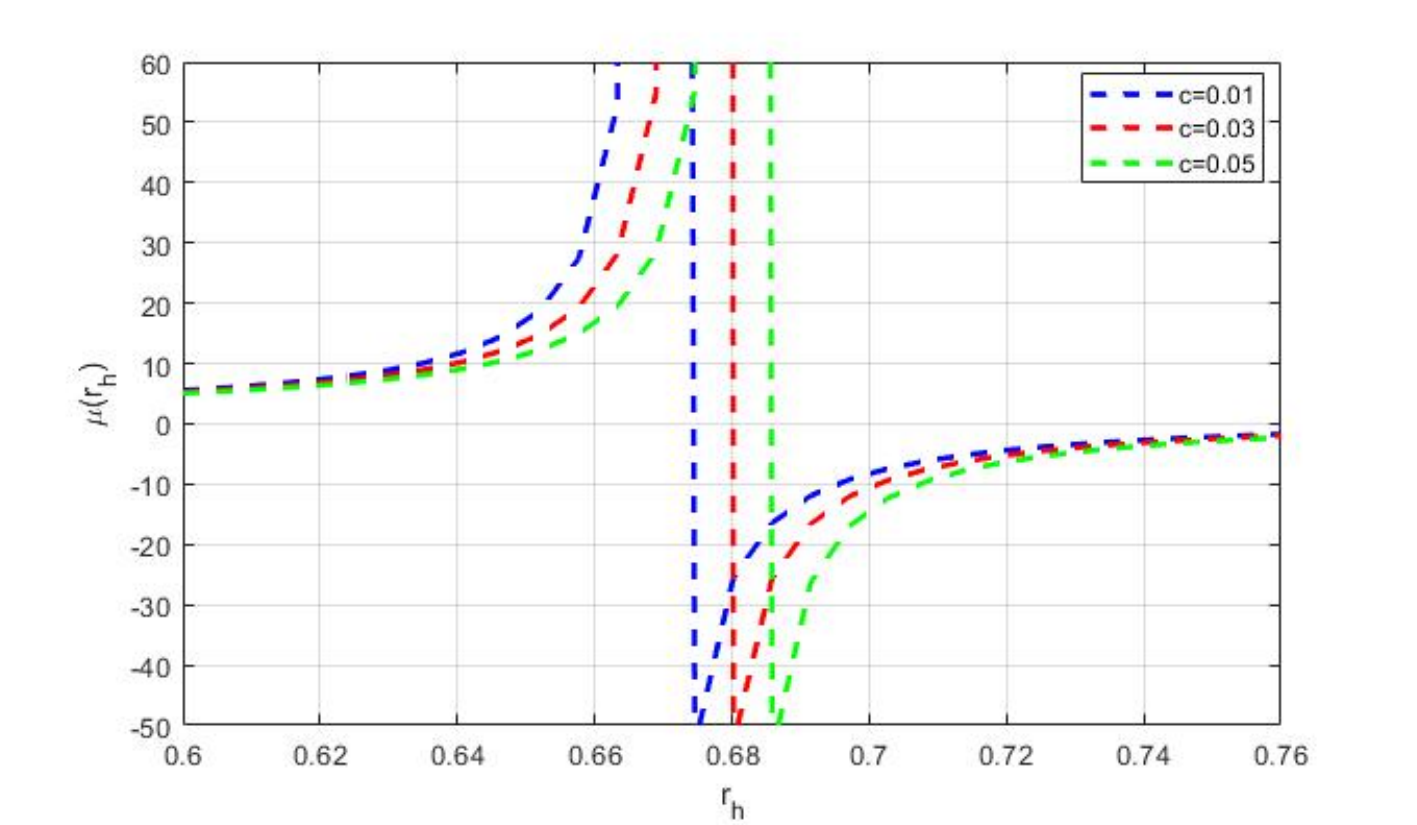}
	
	\vspace{0.3cm}
	
	\includegraphics[width=0.48\textwidth]{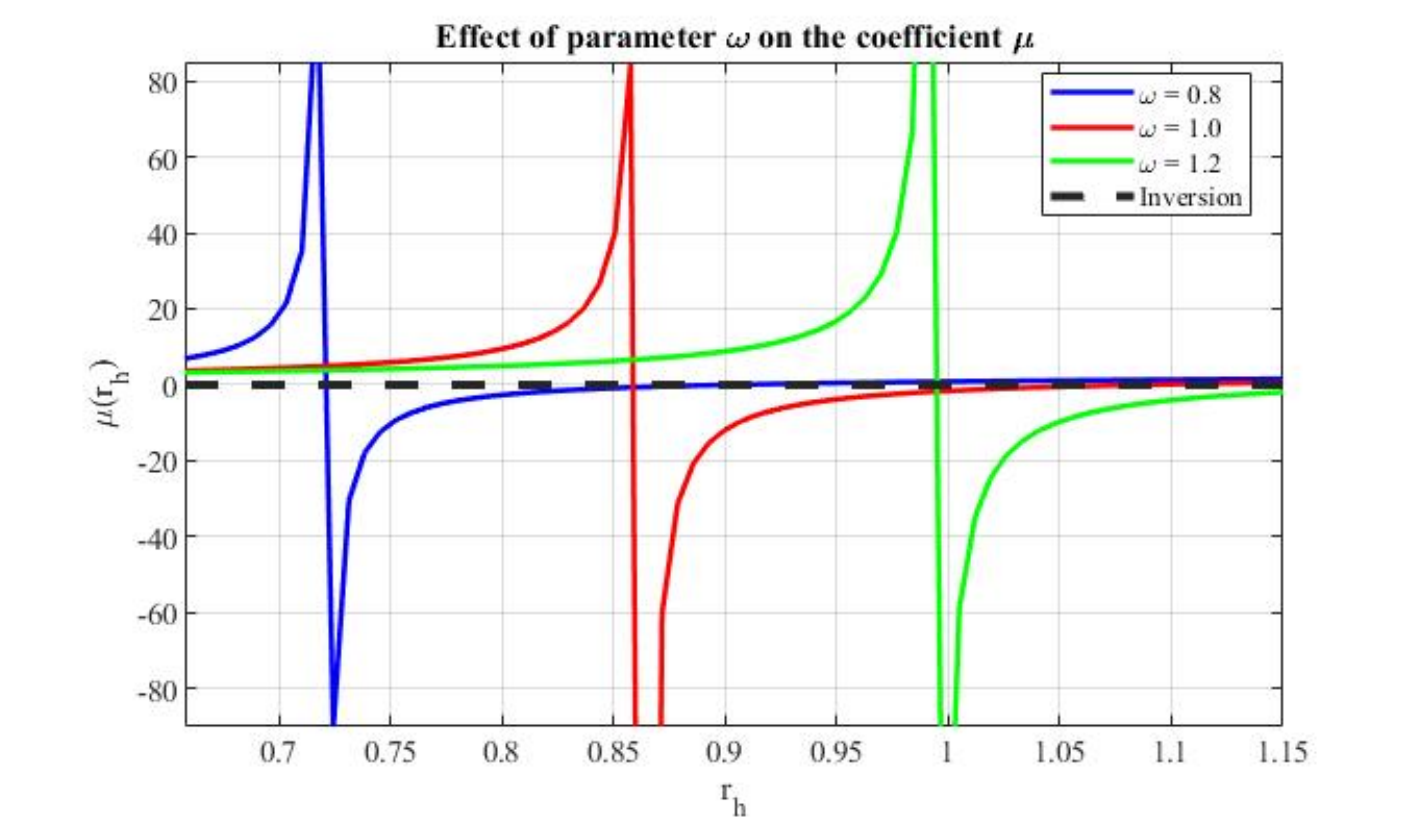}
	\hfill
	\includegraphics[width=0.48\textwidth]{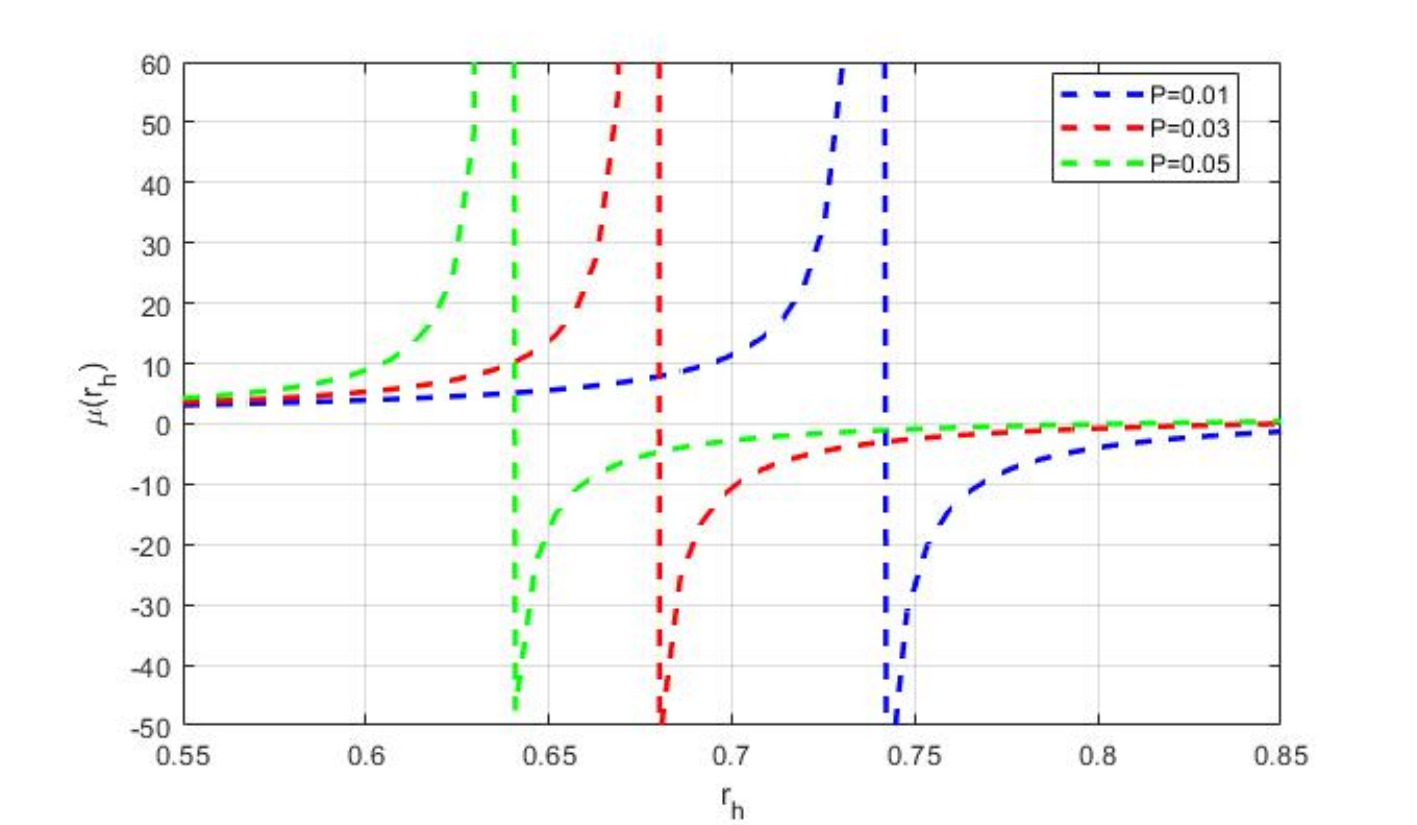}
	
	\caption{ Behavior of the Joule-Thomson coefficient as a function of the horizon radius for different values of the model parameters $\epsilon$, $c$, $\omega$, $Q$, and $P$.}
	\label{F6}
\end{figure}

Fig.~6 displays the JT coefficient $\mu(r_h)$ for different values of $\epsilon$, $c$, $\omega$, and $Q$. The sign of $\mu$ separates the cooling and heating regimes: $\mu>0$ corresponds to cooling, while $\mu<0$ corresponds to heating. The inversion points are determined by the zeros of $\mu$. As shown that, varying $\epsilon$ modifies the width and location of the cooling region. Increasing the quintessence density $c$, changes the position and sharpness of the inversion point.
Variations of the nonlinear parameter $\omega$, modify the electromagnetic contribution and consequently deform the JT profile.
Increasing the dyonic charge $Q$, shifts the inversion radius and changes the strength of the thermal response. These behaviours are consistent with previous studies of JT expansion in charged AdS, quintessence, and nonlinear-electrodynamics black holes [39-49].

The JT coefficient can also be represented as a function of several thermodynamic parameters. Fig.~7 presents the three-dimensional surfaces $\mu(\epsilon,r_h)$, $\mu(Q,r_h)$, and $\mu(P,r_h)$. The surfaces provide a global representation of the cooling and heating regions and their corresponding inversion curves. In particular, less negative values
of $\epsilon$ modify the cooling region, while increasing $Q$ enhances the electromagnetic contribution and shifts the inversion structure. Increasing the pressure changes the position of the inversion curve and modifies the range of horizon radii over which cooling occurs. These results demonstrate that quintessence and nonlinear electrodynamics substantially deform the isenthalpic structure of the black hole.

\begin{figure}[htbp]
	\centering
	\includegraphics[width=0.48\textwidth]{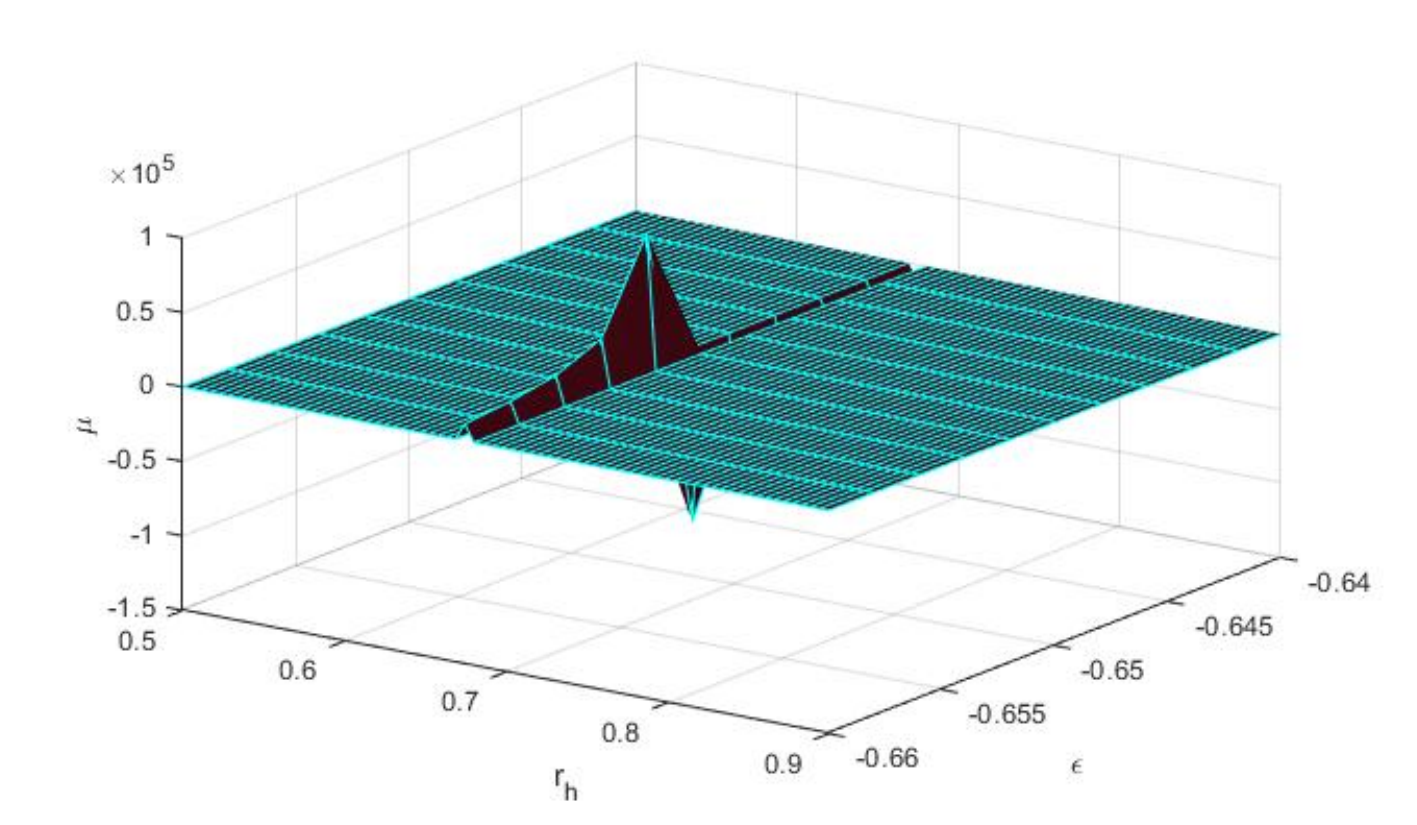}
	\hfill
	\includegraphics[width=0.48\textwidth]{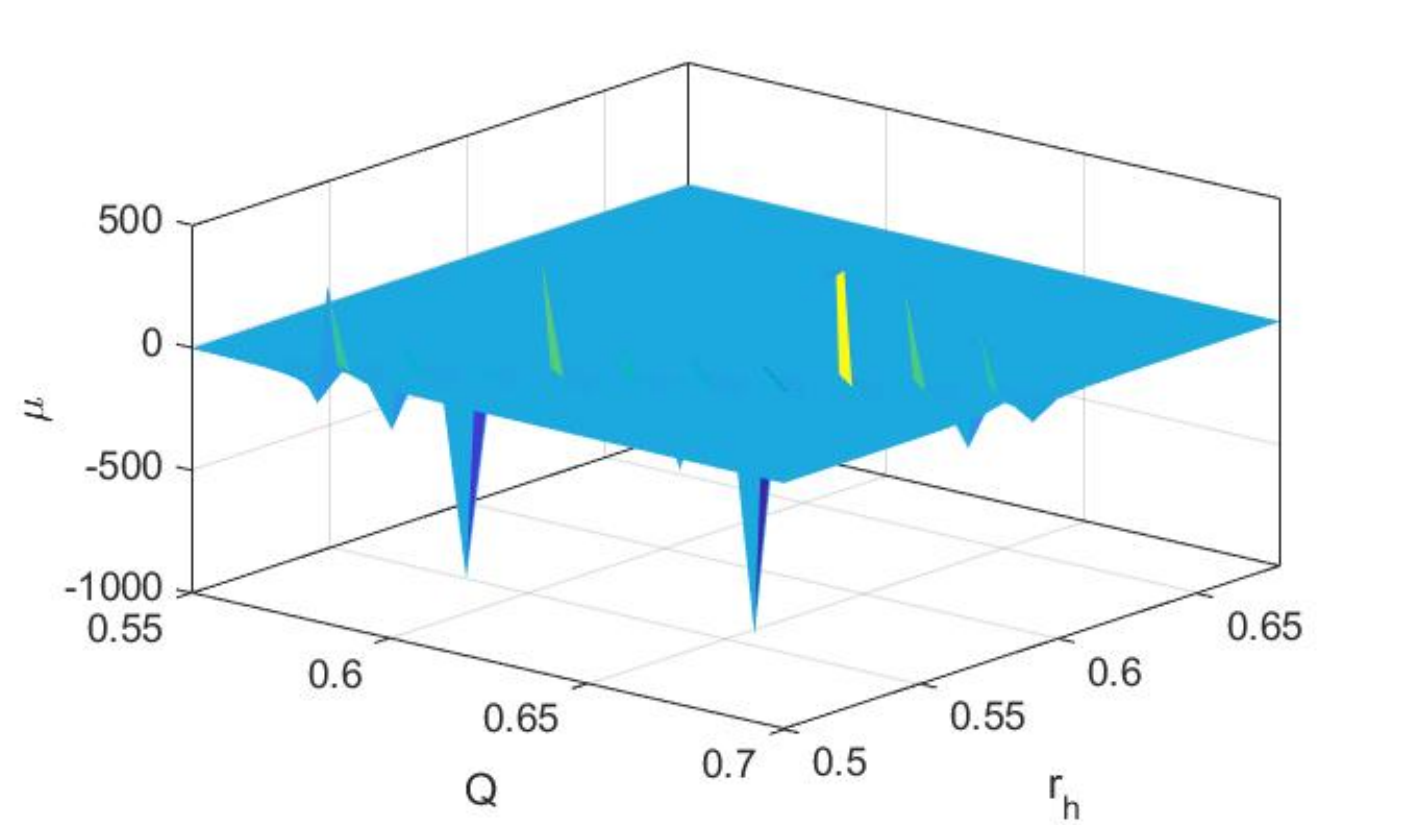}
	
	\vspace{0.3cm}
	
	\includegraphics[width=0.48\textwidth]{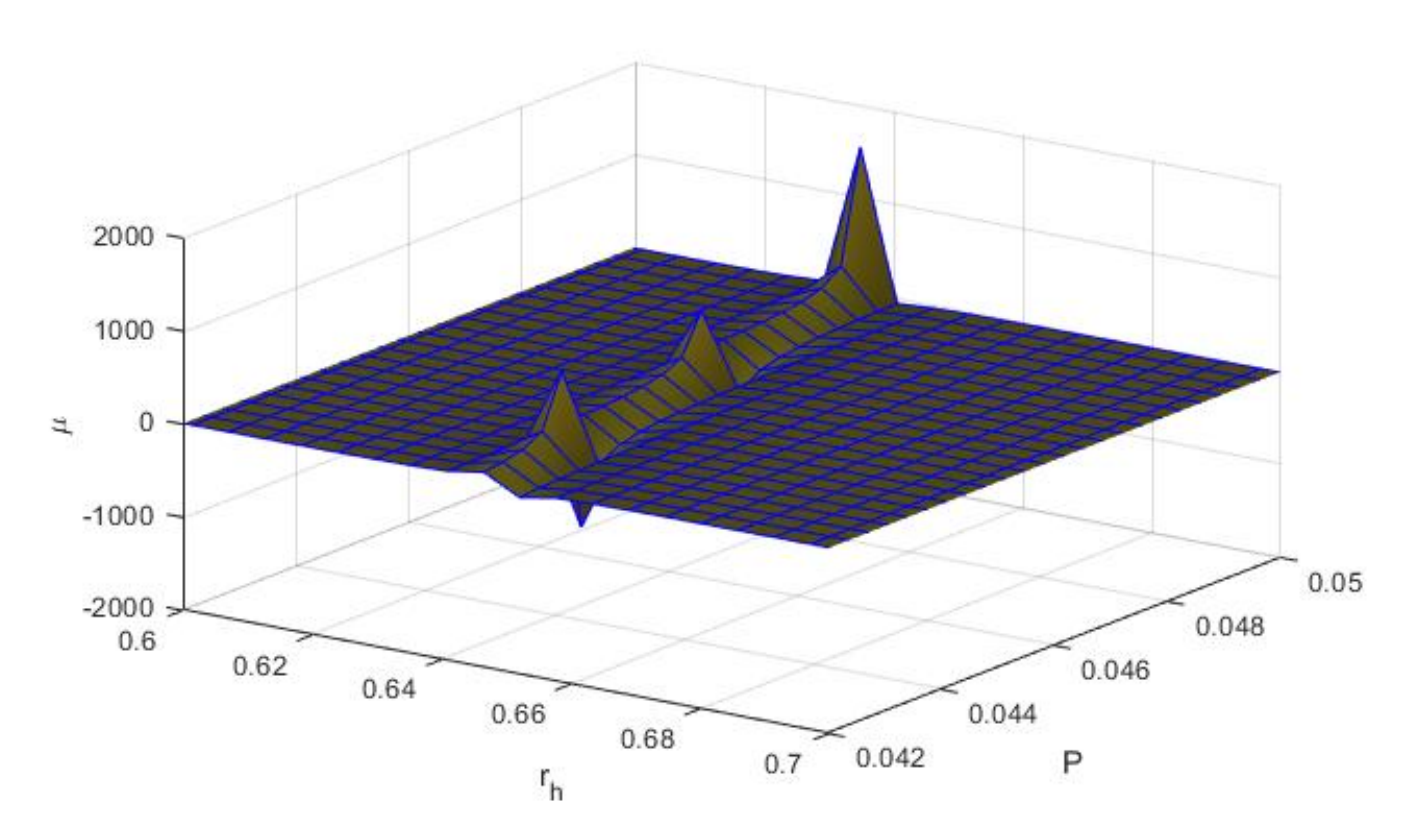}
	
	\caption{Three-dimensional representations of the Joule-Thomson coefficient $\mu$ as functions of the horizon radius and the parameters $\epsilon$, $Q$, and $P$.}
	\label{F7}
\end{figure}
	
The inversion point is defined by
\begin{equation}
\mu=0.
\end{equation}
Using Eq.~(\ref{eq:31}), this condition can equivalently be written as
\begin{equation}
T_i
=
V
\left(\frac{\partial T}{\partial V}\right)_P.
\label{eq:36}
\end{equation}
Since
\begin{equation}
V=\frac{4}{3}\pi r_h^3,
\end{equation}
we have
\begin{equation}
T_i
=
\frac{r_h}{3}
\left(\frac{\partial T}{\partial r_h}\right)_P.
\label{eq:37}
\end{equation}

From Eq.~(30), the derivative of the Hawking temperature with respect to the horizon radius at constant pressure is
\begin{equation}
\left(\frac{\partial T}{\partial r_h}\right)_P
=
-\frac{1}{4\pi r_h^2}
\left[
1-\frac{3\omega^2Q^2}{r_h^2}
-\frac{(3\epsilon+1)(3\epsilon+2)c}
{r_h^{3\epsilon+1}}
-8\pi P r_h^2
\right].
\label{eq:38}
\end{equation}

Consequently, the inversion temperature is
\begin{equation}
T_i
=
\frac{1}{12\pi r_h}
\left[
-1
+\frac{3\omega^2Q^2}{r_h^2}
+\frac{(3\epsilon+1)(3\epsilon+2)c}
{r_h^{3\epsilon+1}}
+8\pi P_i r_h^2
\right],
\label{eq:39}
\end{equation}
where $P_i$ denotes the pressure evaluated on the inversion curve. Equation (\ref{eq:39}) provides the parametric representation of the inversion curve in terms of $r_h$.

In the limit of vanishing quintessence, $c=0$, and in the Maxwell limit $\omega\sim 1$, Eq.~(\ref{eq:39}) reduces to
\begin{equation}
T_i
=
\frac{1}{12\pi r_h}
\left[
-1
+\frac{3Q^2}{r_h^2}
+8\pi P_i r_h^2
\right],
\label{eq:40}
\end{equation}
which is the standard inversion-temperature relation for the
Reissner-Nordstr\"om-AdS black hole [39,40].

The critical inversion temperature corresponds to the minimum temperature of the inversion curve. It is obtained by imposing
\begin{equation}
\left(\frac{\partial T_i}{\partial r_h}\right)_{P_i}=0,
\label{eq:41}
\end{equation}
together with Eq.~(\ref{eq:39}). The resulting critical inversion quantities can then be compared with the critical thermodynamic quantities $(T_c,P_c)$ obtained from the inflection conditions of the equation of state.
This comparison provides a useful characterization of the JT expansion in the vicinity of the black-hole critical point [45,49].

For the Maxwell RN-AdS limit, the inversion temperature obtained from Eq.~(\ref{eq:40}) can be combined with the critical quantities to recover
the well-known relation between the minimum inversion temperature and the critical temperature. In particular, the standard RN-AdS result
\begin{equation}
\frac{T_i^{\,c}}{T_c}=\frac{1}{2}
\label{eq:42}
\end{equation}
is recovered in the corresponding limiting case [39,49]. In the present NED-quintessence model, however, the nonlinear electromagnetic and quintessence contributions modify this ratio, so that it should not be assumed to remain universally equal to $1/2$.

The inversion pressure can be obtained directly from Eq.~(\ref{eq:39}), giving
\begin{equation}
P_i
=
\frac{1}{8\pi r_h^2}
\left[
1-\frac{3\omega^2Q^2}{r_h^2}
-\frac{(3\epsilon+1)(3\epsilon+2)c}
{r_h^{3\epsilon+1}}
+12\pi r_h T_i
\right].
\label{eq:43}
\end{equation}
Equations~(\ref{eq:39}) and (\ref{eq:43}) therefore provide a parametric description of the inversion curve in the $(P_i,T_i)$ plane. They show explicitly how the nonlinear-electrodynamics parameter, the dyonic charge,
and the quintessence field modify the boundary separating the cooling and heating regimes.

The results obtained in this section demonstrate that the JT expansion is highly sensitive to the parameters characterizing the black hole environment. The quintessence field modifies the inversion temperature and the size of the cooling region, while nonlinear electrodynamics and dyonic charge alter the electromagnetic contribution and shift the inversion
curves. These findings extend previous analyses of JT expansion in charged AdS black holes [39], black holes surrounded by dark energy [41], nonlinear charged black holes [43], modified-gravity black holes [45], and NED-AdS black holes [47,48]. The inclusion of the recent analysis of critical
inversion points [49] further motivates the comparison between the inversion structure and the critical behaviour of the present dyonic NED-AdS black hole surrounded by quintessence.

\section{Microscopic Structure via Ruppeiner Thermodynamic Geometry}
\label{sec:ruppeiner}

After analysing the global thermodynamic behaviour and identifying the phase transition structure from the heat capacity and Gibbs free energy, we now investigate the microscopic properties of the black hole using
Ruppeiner thermodynamic geometry. This approach provides a geometric description of thermodynamic fluctuations and allows the effective interactions among the underlying microscopic degrees of freedom to be characterized through the scalar curvature of the thermodynamic state space
[12,50-52]. In particular, the sign and magnitude of the Ruppeiner scalar curvature provide information about the nature and strength of the effective microscopic interactions, while pronounced curvature structures or singularities may be associated with critical behaviour [50-56].

We work in the extended thermodynamic phase space, where the black hole mass is identified with the enthalpy,
\begin{equation}
H\equiv M.
\end{equation}
We choose the two-dimensional thermodynamic state space
\begin{equation}
(x^1,x^2)=(S,P),
\end{equation}
while keeping the dyonic charge $Q$, the nonlinear electrodynamics parameter $\omega$, and the quintessence parameters $c$ and $\epsilon$ fixed. This choice allows us to investigate how the thermodynamic geometry changes with the horizon radius and the thermodynamic pressure [54-61].

Using the relation
\begin{equation}
r_h=\sqrt{\frac{S}{\pi}},
\end{equation}
the enthalpy obtained in Section~2 can be written as
\begin{equation}
M(S,P)
=
\frac{1}{2}
\sqrt{\frac{S}{\pi}}
\left[
1+\frac{\omega^2\pi Q^2}{S}
-c\left(\frac{\pi}{S}\right)^{\frac{3\epsilon+1}{2}}
+\frac{8PS}{3}
\right],
\label{eq:44}
\end{equation}
where
\begin{equation}
Q^2=q_e^2+q_m^2.
\end{equation}

The Weinhold metric is defined by the Hessian of the enthalpy,
\begin{equation}
g^{(W)}_{ij}
=
\frac{\partial^2M}
{\partial x^i\partial x^j},
\end{equation}
and the Ruppeiner metric is related to the Weinhold metric through the conformal transformation
\begin{equation}
g^{(R)}_{ij}
=
\frac{1}{T}g^{(W)}_{ij}
=
\frac{1}{T}
\frac{\partial^2M}
{\partial x^i\partial x^j},
\qquad
(x^1,x^2)=(S,P),
\label{eq:45}
\end{equation}
where the temperature is obtained from
\begin{equation}
T=
\left(\frac{\partial M}{\partial S}\right)_P.
\end{equation}
This construction follows the thermodynamic geometrical framework developed in Refs.~[50-52,54-61].

Differentiating Eq.~(\ref{eq:44}) with respect to the entropy gives
\begin{equation}
T(S,P)
=
\frac{1}{4\sqrt{\pi S}}
\left[
1-\frac{\omega^2\pi Q^2}{S}
-\frac{(3\epsilon+1)c}{2}
\left(\frac{\pi}{S}\right)^{\frac{3\epsilon+1}{2}}
+8PS
\right].
\label{eq:46}
\end{equation}
This expression is algebraically equivalent to the Hawking temperature obtained previously in terms of the horizon radius.

The second derivatives of the enthalpy are the components of the Weinhold metric. The first nonvanishing component is
\begin{equation}
M_{SS}
\equiv
\frac{\partial^2M}{\partial S^2}
=
-\frac{1}{8\sqrt{\pi}S^{3/2}}
\left[
1-\frac{3\omega^2\pi Q^2}{S}
-\frac{(3\epsilon+1)(3\epsilon+3)c}{2}
\left(\frac{\pi}{S}\right)^{\frac{3\epsilon+1}{2}}
-8PS
\right],
\label{eq:47}
\end{equation}
while the mixed derivative is
\begin{equation}
M_{SP}
\equiv
\frac{\partial^2M}{\partial S\partial P}
=
2\sqrt{\frac{S}{\pi}}
=
2r_h.
\label{eq:48}
\end{equation}
Since the enthalpy is linear in the pressure,
\begin{equation}
M_{PP}
\equiv
\frac{\partial^2M}{\partial P^2}
=0.
\label{eq:49}
\end{equation}

Consequently, the Weinhold metric takes the form
\begin{equation}
g^{(W)}_{ij}
=
\begin{pmatrix}
M_{SS} & M_{SP}\\
M_{SP} & 0
\end{pmatrix},
\end{equation}
and the Ruppeiner metric becomes
\begin{equation}
g^{(R)}_{ij}
=
\frac{1}{T}
\begin{pmatrix}
M_{SS} & M_{SP}\\
M_{SP} & 0
\end{pmatrix}.
\end{equation}

For compactness, we introduce
\begin{equation}
A\equiv M_{SS},
\qquad
B\equiv M_{SP},
\qquad
\tau\equiv T.
\end{equation}
The Ruppeiner metric can then be written as
\begin{equation}
g^{(R)}_{ij}
=
\frac{1}{\tau}
\begin{pmatrix}
A & B\\
B & 0
\end{pmatrix}.
\end{equation}
Its determinant is
\begin{equation}
\det g^{(R)}
=
-\frac{B^2}{\tau^2},
\end{equation}
and the inverse metric is
\begin{equation}
\left(g^{(R)}\right)^{-1}
=
\tau
\begin{pmatrix}
0 & B^{-1}\\
B^{-1} & -A/B^2
\end{pmatrix}.
\label{eq:50}
\end{equation}

For a two-dimensional thermodynamic state space, the scalar curvature can be calculated from the metric components. In the coordinates
$(x^1,x^2)=(S,P)$, we use
\begin{equation}
R=
\frac{1}{\sqrt{|g|}}
\left[
\frac{\partial}{\partial x^1}
\left(
\frac{\partial_{x^2}g_{11}
	-\partial_{x^1}g_{12}}
{\sqrt{|g|}}
\right)
+
\frac{\partial}{\partial x^2}
\left(
\frac{\partial_{x^1}g_{22}
	-\partial_{x^2}g_{12}}
{\sqrt{|g|}}
\right)
\right],
\label{eq:51}
\end{equation}
where $g=\det(g_{ij})$. Substituting the particular structure of the Ruppeiner metric, for which $g_{22}=0$, gives the following compact expression:
\begin{equation}
R=
\frac{\tau}
{2B^3}
\left[
B\,\partial_S A
-A\,\partial_S B
+2A\tau\,\partial_P B
-2B
\left(
B\,\partial_P\tau
-\tau\,\partial_P B
\right)
\right].
\label{eq:52}
\end{equation}

Equation~(\ref{eq:52}) provides an exact analytic representation of the Ruppeiner scalar curvature in terms of the thermodynamic quantities $A$, $B$, and $\tau$. Using
\begin{equation}
B=M_{SP}=2r_h,
\end{equation}
together with
\begin{equation}
\tau
=
\frac{1}{4\pi r_h}
\left[
1-\frac{\omega^2Q^2}{r_h^2}
+\frac{3\epsilon c}{r_h^{3\epsilon+1}}
+8\pi P r_h^2
\right],
\end{equation}
and
\begin{equation}
A=M_{SS}
=
\frac{1}{4\pi r_h^2}
\left[
-\frac{1}{2\pi r_h}
+\frac{3\omega^2Q^2}{2\pi r_h^3}
-\frac{3\epsilon(3\epsilon+2)c}
{2\pi r_h^{3\epsilon+2}}
+4Pr_h
\right],
\end{equation}
the curvature can be expressed directly in terms of the horizon radius and the model parameters.

Thus, the Ruppeiner scalar curvature can be written explicitly as
\begin{equation}
R(r_h,P)
=
\frac{
	N(r_h,P;\omega,Q,c,\epsilon)
}{
D(r_h,P;\omega,Q,c,\epsilon)
},
\label{eq:82}
\end{equation}

\begin{equation}
R(r_h,P)
=
\frac{
	4\omega^2Q^2r_h^3
	-2r_h^5
	-9c\epsilon(\epsilon+1)
	r_h^{4-3\epsilon}
}{
\pi r_h^5
\left[
r_h^2-\omega^2Q^2
+3\epsilon c\,r_h^{1-3\epsilon}
+8\pi P r_h^4
\right]
}.
\label{eq:82_explicit}
\end{equation}

where
\begin{align}
N(r_h,P;\omega,Q,c,\epsilon)
&=
4\omega^2Q^2r_h^3
-2r_h^5
-9c\epsilon(\epsilon+1)
r_h^{4-3\epsilon},
\label{eq:N}\\
D(r_h,P;\omega,Q,c,\epsilon)
&=
\pi r_h^5
\left[
r_h^2-\omega^2Q^2
+3\epsilon c\,r_h^{1-3\epsilon}
+8\pi P r_h^4
\right].
\label{eq:D}
\end{align}

The explicit form of Eq.~(\ref{eq:82}) is therefore completely determined by the numerator and denominator given above. In particular, divergences of the Ruppeiner curvature are associated with the zeros of $D(r_h,P;\omega,Q,c,\epsilon)$, provided that the corresponding numerator does not vanish simultaneously.

The sign of the Ruppeiner curvature provides information about the effective microscopic interactions. Following the standard interpretation of Ruppeiner geometry [50-52,58-61], negative values, $R<0$, are associated with predominantly attractive interactions, whereas positive values, $R>0$, indicate predominantly repulsive interactions.
Values close to $R=0$ correspond to weak effective interactions. Strong peaks or divergences of $R$ may signal enhanced thermodynamic correlations and can occur in the vicinity of critical behaviour [55,56,60-65].
\begin{figure}[htbp]
	\centering
	\includegraphics[width=0.48\textwidth]{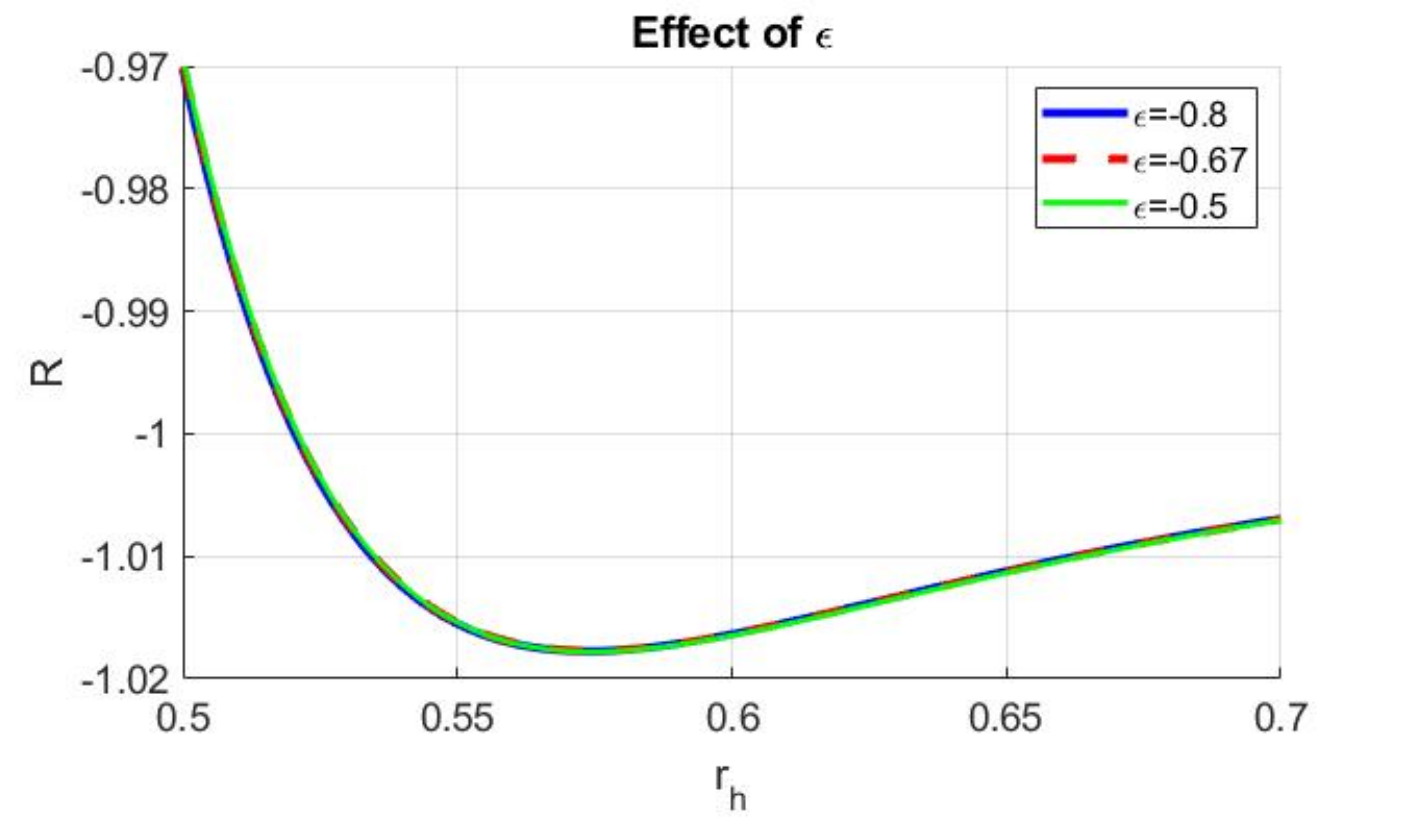}
	\hfill
	\includegraphics[width=0.48\textwidth]{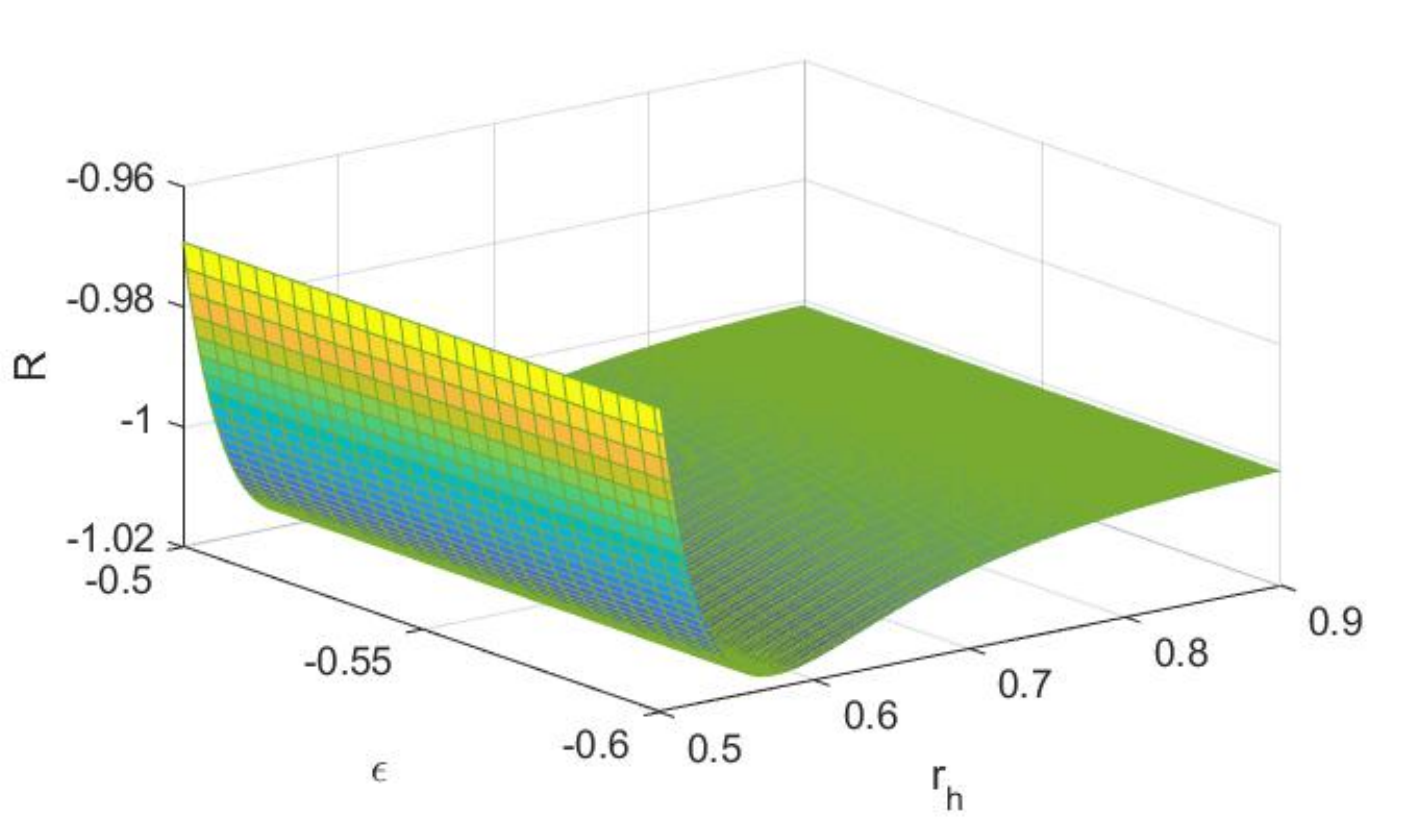}
	\caption{Impact of the parameter state $\epsilon$ on the Ruppeiner curvature $R(r_h)$.}
	\label{F8}
\end{figure}

Fig.~8 shows the dependence of the Ruppeiner curvature on
the quintessence state parameter $\epsilon$. Pronounced peaks and sign changes are observed in $R(r_h)$, indicating strong variations in the effective microscopic interactions. As $\epsilon$ becomes less negative, the curvature peaks are shifted toward larger horizon radii and their magnitude
is reduced. The corresponding three-dimensional surface provides a global representation of the dependence of the curvature on $\epsilon$ and $r_h$.
These results indicate that the equation of state parameter of quintessence has a significant influence on the microscopic correlations of the black hole.

\begin{figure}[!htbp]
	\centering
	\includegraphics[width=0.8\linewidth]{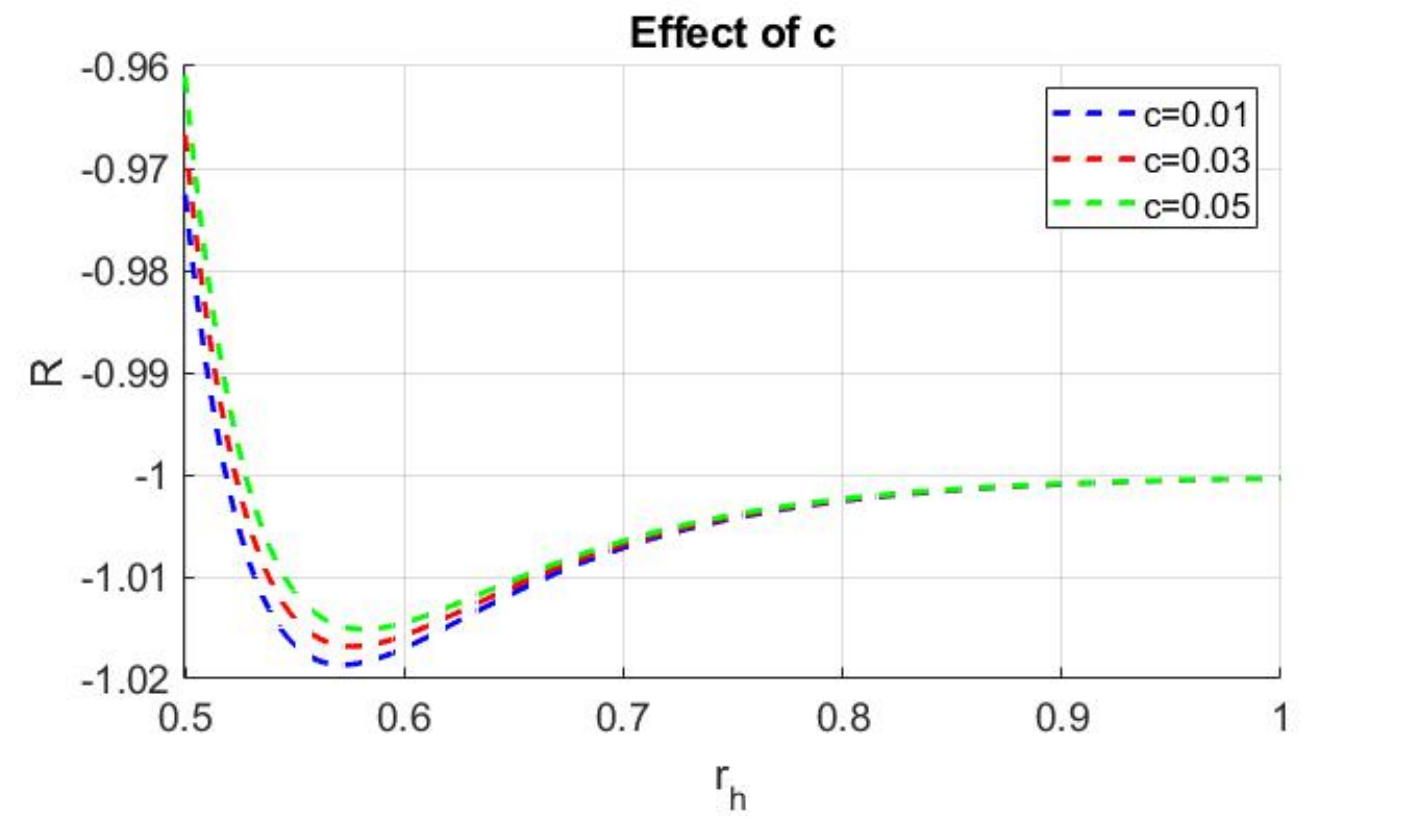}
	\caption{Impact of the quintessence parameter on the Ruppeiner curvature $R(r_h)$.}
	\vspace{-0.2cm}
	\label{F9}
\end{figure}

Fig.~9 investigates the effect of the quintessence density parameter $c$. Increasing $c$ enhances the magnitude of the curvature peaks and shifts their characteristic positions toward larger horizon radii. The sign changes of $R$ persist, indicating alternating regions dominated by attractive and repulsive effective interactions. Thus, a stronger quintessence contribution increases the sensitivity of the microscopic thermodynamic structure.
\begin{figure}[!htbp]
	\centering
	\includegraphics[width=0.8\linewidth]{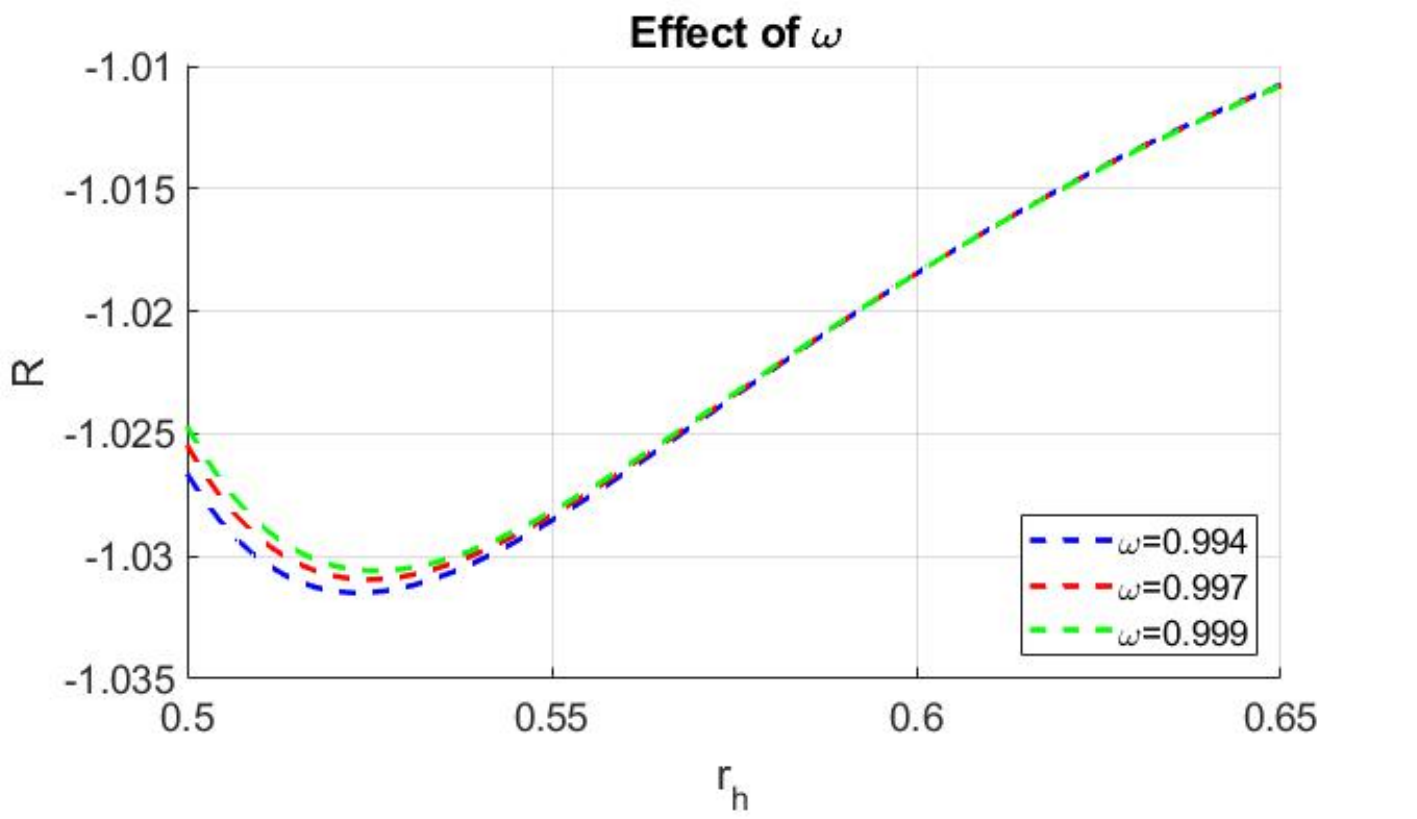}
	\caption{Impact of the NED parameter  $\omega$ on the Ruppeiner curvature $R(r_h)$.}
	\vspace{-0.1cm}
	\label{F10}
\end{figure}

The effect of the nonlinear electrodynamics parameter $\omega$ is shown in Fig.~10. Increasing $\omega$ shifts the characteristic curvature peaks toward larger horizon radii and reduces their magnitude. Consequently, the curvature becomes smoother as the system approaches the Maxwell limit
$\omega \sim 1$. The persistence of sign changes in $R$ indicates that both attractive and repulsive effective interactions can occur. These results demonstrate that nonlinear electrodynamics provides an important modification of the microscopic structure of the black hole, in agreement with previous studies of NED black-hole thermodynamic geometry
[47,48,64].

\begin{figure}[!htbp]
	\centering
	\includegraphics[width=0.8\linewidth]{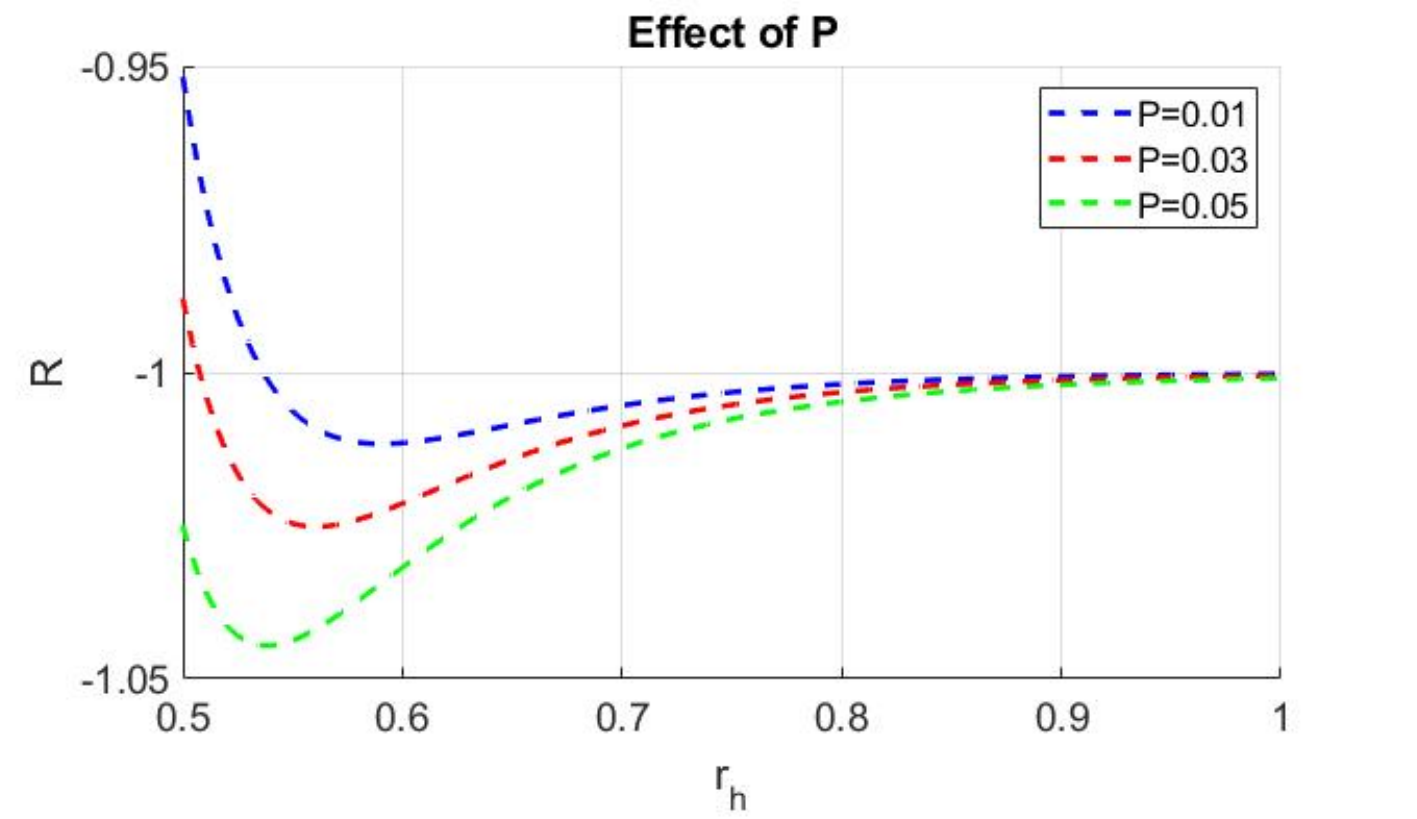}
	\caption{Influence of the thermodynamic pressure P(0.01 0.03 0.05) on the Ruppeiner curvature $R(r_h)$.}
	\vspace{-0.1cm}
	\label{F11}
\end{figure}

Fig.~11 illustrates the effect of the thermodynamic pressure $P$ on the Ruppeiner curvature. Increasing the pressure shifts the characteristic curvature peaks toward smaller horizon radii and reduces their magnitude.
This behaviour indicates a weakening and narrowing of the regions with strong thermodynamic correlations. The sign changes of $R$ remain present, showing that both attractive and repulsive effective interactions persist.

\begin{figure}[!htbp]
	\centering
	\includegraphics[width=0.8\linewidth]{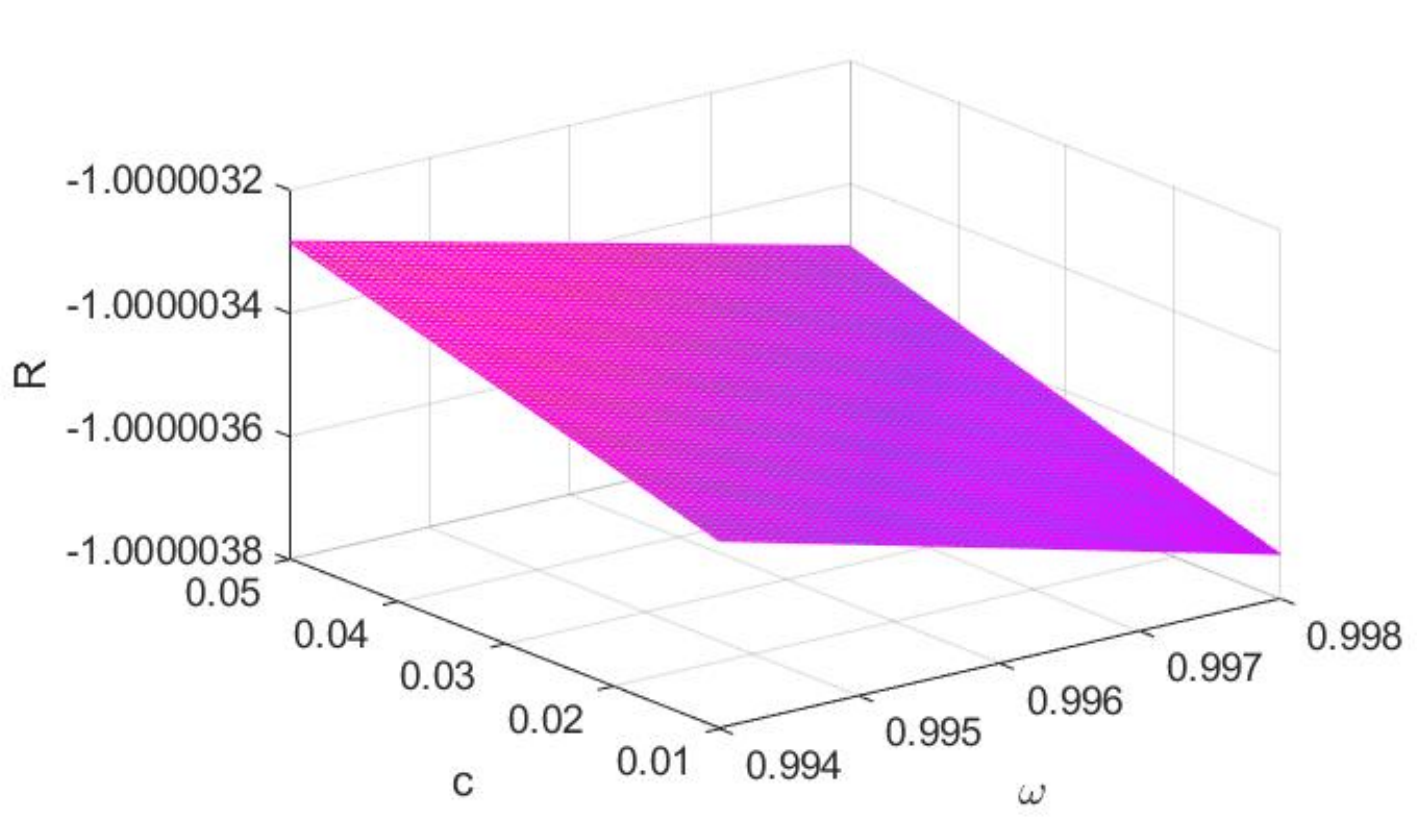}
	\caption{3D visualization of the dependence of c and $\omega$ on R.}
	\vspace{-0.1cm}
	\label{F12}
\end{figure}

Finally, Fig.~12 displays the combined dependence of the Ruppeiner curvature on the quintessence density parameter $c$ and the nonlinear-electrodynamics parameter $\omega$. Increasing $c$ enhances the magnitude of the curvature and enlarges the regions of strong thermodynamic
correlations, whereas increasing $\omega$ shifts the curvature extrema and reduces their intensity. The competition between these two parameters therefore produces a nontrivial modification of the microscopic structure.
In particular, the approach to the Maxwell limit tends to smooth the curvature profile, while a stronger quintessence contribution enhances the magnitude of the thermodynamic correlations.
The Ruppeiner analysis complements the macroscopic thermodynamic and Joule-Thomson analyses developed in the preceding sections. The curvature encodes information about effective microscopic interactions that cannot be obtained directly from the equation of state or heat capacity alone.
Therefore, the simultaneous analysis of the sign, magnitude, and parameter dependence of $R$ provides a geometrical perspective on how nonlinear electrodynamics, dyonic charge, quintessence, and pressure influence the microscopic structure of the black hole [58-65].

\section{Conclusion}
In this work, we have investigated the thermodynamic properties, Joule-Thomson expansion, and microscopic structure of a dyonic Reissner-Nordstr\"om-AdS black hole in nonlinear electrodynamics surrounded by quintessence. The black hole thermodynamic quantities were derived in the extended phase space, including the Hawking temperature (20), equation of state (26), heat capacity (30), and Gibbs free energy (32).
The temperature profiles in Fig.~2 and the equation of state in Fig.~3 reveal a Van der Waals-like phase structure. The critical conditions (33) lead to the critical quantities given by Eqs.~(34)-(36). The heat capacity analysis in Fig.~4 shows divergences associated with second-order phase transition behaviour, while the Gibbs free energy in Fig.~5 confirms the existence of a nontrivial phase structure. The quintessence parameters $(c,\epsilon)$, the nonlinear electrodynamics parameter $\omega$, and the dyonic charge $Q$ significantly modify the critical behaviour.
We have also analysed the Joule-Thomson expansion through the coefficient $\mu$ given in Eqs.~(39)-(49). The results in Figs.~6 and 7 show distinct cooling and heating regions separated by inversion points. Quintessence generally lowers the inversion temperature, whereas the nonlinear electrodynamics parameter and the dyonic charge shift the inversion curves and modify the cooling region. The corresponding inversion relations are obtained from Eqs.~(50)-(59).
Finally, the microscopic structure was investigated using Ruppeiner geometry. The thermodynamic metric and scalar curvature were derived in Eqs.~(68)-(86). As shown in Figs.~8-12, the Ruppeiner curvature is strongly affected by $\epsilon$, $c$, $\omega$, and $P$, exhibiting pronounced peaks and sign changes near critical regions. Positive and negative values of $R$ are interpreted as predominantly repulsive and attractive effective microscopic interactions, respectively. Increasing $c$ enhances the magnitude of the curvature, whereas increasing $\epsilon$ and $\omega$ generally smooths the curvature structure.
Overall, our results demonstrate that nonlinear electrodynamics, dyonic charge, quintessence, and AdS pressure jointly reshape the thermodynamic phase structure, Joule-Thomson inversion behaviour, and effective microscopic interactions of the black hole. The combined thermodynamic and geometrical analysis therefore provides a useful framework for exploring the interplay between nonlinear electromagnetic effects and dark energy environments in AdS black hole systems.
These findings may further illuminate the holographic interpretation of AdS black holes and their quantum thermodynamic duals.


\begin{thebibliography}{99}
	
	\bibitem{Riess1998} A. G. Riess et al., “Observational Evidence from Supernovae for an Accelerating Universe and a Cosmological Constant,” \textit{Astron. J.} \textbf{116}, 1009 (1998).
	
	\bibitem{Perlmutter1999} S. Perlmutter et al., “Measurements of $\Omega$ and $\Lambda$ from 42 High-Redshift Supernovae,” \textit{Astrophys. J.} \textbf{517}, 565 (1999).
	
	\bibitem{Planck2020} N. Aghanim et al. (Planck Collaboration), “Planck 2018 results. VI. Cosmological parameters,” \textit{Astron. Astrophys.} \textbf{641}, A6 (2020).
	
	\bibitem{Bekenstein1973} J. D. Bekenstein, “Black holes and entropy,” \textit{Phys. Rev. D} \textbf{7}, 2333 (1973).
	
	\bibitem{Hawking1975} S. W. Hawking, “Particle creation by black holes,” \textit{Commun. Math. Phys.} \textbf{43}, 199 (1975).
	
	\bibitem{Bardeen1973} J. M. Bardeen, B. Carter, and S. W. Hawking, “The four laws of black hole mechanics,” \textit{Commun. Math. Phys.} \textbf{31}, 161 (1973).
	
	\bibitem{Kubiznak2012} D. Kubizňák and R. B. Mann, “P–V criticality of charged AdS black holes,” \textit{JHEP} \textbf{1207}, 033 (2012).
	
	\bibitem{Chamblin1999a} A. Chamblin, R. Emparan, C. V. Johnson, and R. C. Myers, “Charged AdS black holes and catastrophic holography,” \textit{Phys. Rev. D} \textbf{60}, 064018 (1999).
	
	\bibitem{Chamblin1999b} A. Chamblin, R. Emparan, C. V. Johnson, and R. C. Myers, “Holography, thermodynamics and fluctuations of charged AdS black holes,” \textit{Phys. Rev. D} \textbf{60}, 104026 (1999).
	
	\bibitem{Kastor2009} D. Kastor, S. Ray, and J. Traschen, “Enthalpy and the mechanics of AdS black holes,” \textit{Class. Quantum Grav.} \textbf{26}, 195011 (2009).
	
	\bibitem{Kubiznak2017} D. Kubizňák, R. B. Mann, and M. Teo, “Black hole chemistry: thermodynamics with Lambda,” \textit{Class. Quantum Grav.} \textbf{34}, 063001 (2017).
	
	\bibitem{Wei2019} S.-W. Wei and Y.-X. Liu, “Microstructure and Ruppeiner geometry of charged AdS black holes,” \textit{Phys. Rev. D} \textbf{100}, 086007 (2019).
	
	\bibitem{Ortin2004} T. Ortín, \textit{Gravity and Strings}, Cambridge University Press (2004).
	
	\bibitem{Rasheed1997} D. Rasheed, “Nonlinear electrodynamics: zeroth and first laws of black hole mechanics,” arXiv:hep-th/9702087 (1997).
	
	\bibitem{Maldacena1999} J. Maldacena, “The Large-N limit of superconformal field theories and supergravity,” \textit{Int. J. Theor. Phys.} \textbf{38}, 1113 (1999).
	
	\bibitem{Amirabi2021}
	Z.~Amirabi and S.~H. Mazharimousavi,
	\newblock {\em Black-hole solution in nonlinear electrodynamics with the maximum allowable symmetries},
	\newblock Eur. Phys. J. C {\bf 81}, 207 (2021).
	
	\bibitem{Born1934} M. Born and L. Infeld, “Foundations of the new field theory,” \textit{Proc. R. Soc. Lond. A} \textbf{144}, 425 (1934).
	
	\bibitem{AyonBeato1998} E. Ayón-Beato and A. García, “Regular black hole in general relativity coupled to nonlinear electrodynamics,” \textit{Phys. Rev. Lett.} \textbf{80}, 5056 (1998).
	
	\bibitem{Bronnikov2001} K. A. Bronnikov, “Regular magnetic black holes and monopoles from nonlinear electrodynamics,” \textit{Phys. Rev. D} \textbf{63}, 044005 (2001).
	
	\bibitem{Hendi2012} S. H. Hendi, “Asymptotic Reissner–Nordström black holes,” \textit{Ann. Phys.} \textbf{333}, 282 (2012).
	
	\bibitem{Hendi2016} S. H. Hendi, B. Eslam Panah, and S. Panahiyan, “Geometrical thermodynamics and P–V criticality of black holes with a nonlinear source,” \textit{Int. J. Mod. Phys. D} \textbf{25}, 1650010 (2016).
	
	\bibitem{Breton2015} N. Breton and R. Garcia-Salcedo, \textit{Nonlinear Electrodynamics and Black Holes}, Springer (2015).
	
	\bibitem{Kruglov2021} S. I. Kruglov, “Nonlinear electrodynamics and magnetic black holes,” \textit{Ann. Phys.} \textbf{427}, 168425 (2021).
	
	\bibitem{Ali2022} M. S. Ali and S. G. Ghosh, “Thermodynamics of nonlinear electrodynamics black holes in AdS space,” \textit{Phys. Lett. B} \textbf{835}, 137567 (2022).
	
	\bibitem{Hu2023} Y. Hu and H. Xu, “Phase transition and Joule–Thomson expansion of NED-AdS black holes,” \textit{Eur. Phys. J. C} \textbf{83}, 542 (2023).
	
	\bibitem{Zhang2024} J. Zhang, X. Liu, and S.-W. Wei, “Joule–Thomson expansion of nonlinear AdS black holes and microstructure analysis,” \textit{Phys. Rev. D} \textbf{109}, 064001 (2024).
	
	\bibitem{Mahapatra2023} S. Mahapatra, “Thermodynamic phase transitions in nonlinear charged AdS black holes,” \textit{JHEP} \textbf{2308}, 051 (2023).
	
	\bibitem{Caldwell1998} R. R. Caldwell, R. Dave, and P. J. Steinhardt, “Cosmological imprint of an energy component with general equation of state,” \textit{Phys. Rev. Lett.} \textbf{80}, 1582 (1998).
	
	\bibitem{Kiselev2003} V. V. Kiselev, “Quintessence and black holes,” \textit{Class. Quantum Grav.} \textbf{20}, 1187 (2003).
	
	\bibitem{Chen2008} S. Chen, B. Wang, and R. Su, “Hawking radiation in a dS black hole surrounded by quintessence,” \textit{Phys. Rev. D} \textbf{77}, 124011 (2008).
	
	\bibitem{Azreg2014} M. Azreg-Aïnou, “Thermodynamics of nonlinear electrodynamics black holes with quintessence,” \textit{Eur. Phys. J. C} \textbf{74}, 2865 (2014).
	
	\bibitem{Kofane2018}
	M.~Saleh, B.~B. Thomas, and T.~C. Kofane,
	\newblock {\em Thermodynamics and phase transition of the Reissner–Nordström black hole surrounded by quintessence},
	\newblock Gen. Relativ. Gravit. {\bf 44}, 2181 (2012).
	
	\bibitem{Pradhan2020} P. Pradhan, “Thermodynamic geometry of black holes surrounded by quintessence,” \textit{Mod. Phys. Lett. A} \textbf{35}, 2050128 (2020).
	
	\bibitem{Thomas2020}
	J.~Rodrigue, M.~Saleh, B.~B. Thomas, and T.~C. Kofane,
	\newblock {\em Thermodynamic phase transition and global stability of the regular Hayward black hole surrounded by quintessence},
	\newblock Mod. Phys. Lett. A {\bf 35}, 2050136 (2020).
	
	\bibitem{Ghosh2022} S. G. Ghosh and A. Banerjee, “Thermodynamics of black holes in quintessence,” \textit{Eur. Phys. J. C} \textbf{82}, 1050 (2022).
	
	\bibitem{Ahmed2021} M. Ahmed and U. Camci, “Black hole thermodynamics in quintessence and PFDM backgrounds,” \textit{Ann. Phys.} \textbf{429}, 168482 (2021).
	
	\bibitem{Singh2023} D. V. Singh, “Quintessence dark energy and black hole phase transitions,” \textit{Chin. Phys. C} \textbf{47}, 045103 (2023).
	
	\bibitem{Wei2024} S.-W. Wei and Y.-X. Liu, “Quintessence, phase transition and microstructure of AdS black holes,” \textit{Phys. Rev. D} \textbf{109}, 084041 (2024).
	
	\bibitem{Okcu2017} O. Ökcü and E. Aydıner, “Joule–Thomson expansion of the charged AdS black holes,” \textit{Eur. Phys. J. C} \textbf{77}, 24 (2017).
	
	\bibitem{Okcu2018} O. Ökcü and E. Aydıner, “Joule–Thomson expansion of Kerr–AdS black holes,” \textit{Eur. Phys. J. C} \textbf{78}, 123 (2018).
	
	\bibitem{Mo2018} J.-X. Mo, G.-Q. Li, and X.-B. Xu, “Effects of dark energy on Joule–Thomson expansion of AdS black holes,” \textit{Phys. Rev. D} \textbf{98}, 124032 (2018).
	
	\bibitem{Cisterna2018} A. Cisterna, S. Q. Hu, and S. E. Jorás, “Joule–Thomson expansion in Gauss–Bonnet black holes,” \textit{Phys. Lett. B} \textbf{779}, 1 (2018).
	
	\bibitem{Ahmed2020} M. Ahmed, S. H. Hendi, and A. Sheykhi, “Joule–Thomson expansion for nonlinear charged black holes,” \textit{Ann. Phys.} \textbf{417}, 168116 (2020).
	
	\bibitem{Haldar2019} A. Haldar and R. Biswas, “Joule–Thomson expansion for charged Lovelock black holes,” \textit{Gen. Relativ. Gravit.} \textbf{51}, 72 (2019).
	
	\bibitem{Zhang2021} J. Zhang, Y. Liu, and S.-W. Wei, “Joule–Thomson expansion and microstructure of modified gravity black holes,” \textit{Phys. Rev. D} \textbf{104}, 084086 (2021).
	
	\bibitem{Mo2022} J.-X. Mo and W.-B. Liu, “Joule–Thomson expansion of AdS black holes surrounded by quintessence,” \textit{Eur. Phys. J. C} \textbf{82}, 931 (2022).
	
	\bibitem{Ali2023} M. S. Ali, S. G. Ghosh, and B. P. Singh, “Thermodynamic geometry and Joule–Thomson expansion of NED black holes,” \textit{Phys. Rev. D} \textbf{108}, 024006 (2023).
	
	\bibitem{Hendi2024} S. H. Hendi and M. Panahiyan, “Joule–Thomson expansion and phase transition in nonlinear electrodynamics AdS black holes,” \textit{Eur. Phys. J. C} \textbf{84}, 404 (2024).
	
	\bibitem{Wei2025} S.-W. Wei, Y.-X. Liu, and R. B. Mann, “Critical inversion points and phase transitions in black hole Joule–Thomson expansion,” \textit{JHEP} \textbf{2503}, 065 (2025).
	
	\bibitem{Ruppeiner1979} G. Ruppeiner, “Thermodynamics: a Riemannian geometric model,” \textit{Phys. Rev. A} \textbf{20}, 1608 (1979).
	
	\bibitem{Ruppeiner1995} G. Ruppeiner, “Riemannian geometry in thermodynamic fluctuation theory,” \textit{Rev. Mod. Phys.} \textbf{67}, 605 (1995).
	
	\bibitem{Ruppeiner2014} G. Ruppeiner, “Thermodynamic curvature and black holes,” \textit{Springer Lect. Notes Phys.} \textbf{905}, 179 (2014).
	
	\bibitem{Quevedo2008} H. Quevedo, “Geometrothermodynamics,” \textit{Gen. Relativ. Gravit.} \textbf{40}, 971 (2008).
	
	\bibitem{Sarkar2006} T. Sarkar, G. Sengupta, and B. N. Tiwari, “On the thermodynamic geometry of black holes,” \textit{JHEP} \textbf{0611}, 015 (2006).
	
	\bibitem{Sahay2010} A. Sahay, T. Sarkar, and G. Sengupta, “Thermodynamic geometry and phase transitions in Kerr–Newman–AdS black holes,” \textit{JHEP} \textbf{1004}, 118 (2010).
	
	\bibitem{Mansoori2016} S. A. H. Mansoori and B. Mirza, “Correspondence of phase transition points and singularities of thermodynamic geometry,” \textit{Eur. Phys. J. C} \textbf{74}, 2681 (2016).
	
	\bibitem{Aman2003} J. E. Åman, I. Bengtsson, and N. Pidokrajt, “Geometry of black hole thermodynamics,” \textit{Gen. Relativ. Gravit.} \textbf{35}, 1733 (2003).
	
	\bibitem{Mirza2009} B. Mirza and M. Zamaninasab, “Ruppeiner geometry of RN black holes,” \textit{JHEP} \textbf{06}, 059 (2009).
	
	\bibitem{Miao2018} Y.-G. Miao and Z.-M. Xu, “On thermal molecular potential and Ruppeiner geometry of RN–AdS black holes,” \textit{Phys. Rev. D} \textbf{98}, 044001 (2018).
	
	\bibitem{Mo2019} J.-X. Mo, “Ruppeiner geometry and phase transitions in AdS black holes,” \textit{Eur. Phys. J. C} \textbf{79}, 9 (2019).
	
	\bibitem{Wei2020} S.-W. Wei and Y.-X. Liu, “Insight into microscopic structure of RN–AdS black hole from thermodynamic curvature,” \textit{Phys. Rev. D} \textbf{101}, 104018 (2020).
	
	\bibitem{Dehyadegari2022} A. Dehyadegari and B. Mirza, “Ruppeiner geometry of topological black holes,” \textit{Eur. Phys. J. C} \textbf{82}, 142 (2022).
	
	\bibitem{Pradhan2023b} P. Pradhan, “Ruppeiner geometry and stability of AdS black holes surrounded by quintessence,” \textit{Mod. Phys. Lett. A} \textbf{38}, 2350129 (2023).
	
	\bibitem{Li2024} J. Li and Y. Chen, “Thermodynamic geometry and phase structure of NED-AdS black holes,” \textit{Eur. Phys. J. C} \textbf{84}, 390 (2024).
	
	\bibitem{Hu2025} H. Hu, Y. Xu, and S. Zhang, “Thermodynamic curvature and microscopic interactions of black holes with dark energy,” \textit{Phys. Rev. D} \textbf{111}, 044022 (2025).	
\end{thebibliography}

\end{document}